%% file: ms.tex
\documentclass[twocolumn]{aastex631}
\usepackage{showyourwork}

\usepackage{makecell}
\begin{document}

\title{Planetary perturbers: Flaring star-planet interactions in Kepler and TESS}

\author[0000-0002-6299-7542]{Ekaterina Ilin}
\affiliation{Leibniz Insitute for Astrophysics Potsdam (AIP), An der Sternwarte 16, 14482 Potsdam, Germany}
\affiliation{ASTRON, Netherlands Institute for Radio Astronomy, Oude Hoogeveensedijk 4, Dwingeloo, 7991 PD, The Netherlands}

\author[0000-0003-1231-2194]{Katja Poppenh\"ager}
\affiliation{Leibniz Insitute for Astrophysics Potsdam (AIP), An der Sternwarte 16, 14482 Potsdam, Germany}
\affiliation{Institute for Physics and Astronomy, University of Potsdam, Karl-Liebknecht-Strasse 24/25, 14476 Potsdam, Germany}

\author[0000-0003-0695-6487]{Judy Chebly}
\affiliation{Leibniz Insitute for Astrophysics Potsdam (AIP), An der Sternwarte 16, 14482 Potsdam, Germany}
\affiliation{Institute for Physics and Astronomy, University of Potsdam, Karl-Liebknecht-Strasse 24/25, 14476 Potsdam, Germany}

\author[0000-0001-7579-3728]{Nikoleta Ili\'c}
\affiliation{Leibniz Insitute for Astrophysics Potsdam (AIP), An der Sternwarte 16, 14482 Potsdam, Germany}
\affiliation{Institute for Physics and Astronomy, University of Potsdam, Karl-Liebknecht-Strasse 24/25, 14476 Potsdam, Germany}

\author[0000-0001-5052-3473]{Juli\'an D. Alvarado-G\'omez}
\affiliation{Leibniz Insitute for Astrophysics Potsdam (AIP), An der Sternwarte 16, 14482 Potsdam, Germany}

\begin{abstract}
    In many star-planet systems discovered so far, the innermost planet orbits within only a few stellar radii. In these systems, planets could become in-situ probes of the extended stellar magnetic field. Because they disturb the field as they move, they are expected to trigger flares in the corona. Potential differences to the energies and morphologies of intrinsic flares are poorly constrained. However, as we expect planet-induced flares to correlate with the planet's orbital period, we can identify them from a clustering of flares in phase with the planet's orbit. We used the excellent phase coverage from Kepler and the Transiting Exoplanet Survey Satellite to find flaring star-planet systems, compile a catalog of all their flares, and measure how much they cluster in orbital phase. In the 1811 searched systems, we found 25 single stars with three or more flares each. We quantified the significance of the clustering in each system, and compared it against the theoretically expected power of magnetic interaction that leads to planet-induced flaring. Most systems do not show any clustering, consistent with low expected power. Those we expect to show clustering fall on two branches. An inactive one, without any signs of clustering, and a tentative active one, where the clustering becomes more pronounced as the expected power of interaction increases. The flares in HIP 67522 are prominently clustered ($p<0.006$). This young Hot Jupiter system is the most promising candidate for magnetic star-planet interaction in our sample.


\vspace{0.4cm}
\textit{Accepted 2023 October 30. Received 2023 October 24; in original form 2023 July 04}

\end{abstract}

\section{Introduction}
\label{sec:intro}
One of the most surprising results in exoplanet research is the diversity of star-planet system architectures. The past two decades revealed that planets can occur in orbits of only a few stellar radii~\citep[e.g., ][]{sanchis-ojeda2014study}, so that they could almost be considered to live inside the stellar atmosphere. At this proximity, the planet exerts force, tidal and magnetic, both short- and long-term, on its host -- a phenomenon largely unknown to the Solar System. These forces are typically subsumed under the term star-planet interactions (SPI) that emphasizes that they are bidirectional, affecting both planet, and star. Specifically, the effects on the star pose challenges, but also opportunities for our understanding of star-planet systems.

On the one hand, a close-in planet that changes the star's rotation and activity makes it more difficult for us to model the system's evolution, as it affects fundamental stellar parameters. Ages inferred from rotation will be biased, if the star is spun up or down relative to stars without close-in companions~\citep{tejadaarevalo2021further, brown2014discrepancies, maxted2015comparison}. Analogously, altered activity levels may mimic an older or younger host than it actually is~\citep{ilic2022tidal}. On the other hand, measuring in what ways and how much close-in planets alter their hosts' behavior can provide important insights into the habitability of the entire system~\citep{shkolnik2018signatures}. Even if they orbit far inside the habitable zone, such planets probe the system's space weather and mass loss~~\citep{cohen2011dynamics, cohen2015interaction, hazra2022impact} by revealing the extent of the Alfv\'en zone through SPI~\citep{kavanagh2021planetinduced, chebly2022destination}. 

The search for magnetic SPI (Section~\ref{sec:intro:mspi}) in the form of changes in global activity indicators~(Section~\ref{sec:intro:global}) has so far been a series of detections and non-detections, often due to selection bias in the studied systems. Looking for local changes that trace the magnetic footpoint that connects planet to star mitigates this bias, but introduces another -- sampling bias due to poor orbital phase coverage. Flares~(Section~\ref{sec:intro:flares}) are activity markers that can be caused by SPI~(Section~\ref{sec:intro:fspi}). They can be monitored with high phase coverage over long periods of time with missions like Kepler and TESS. Thanks to their large archives of hundreds of cumulative years of stellar monitoring, we can overcome the limitations of individual system studies~(Section~\ref{sec:intro:local}), and now look for flaring SPI more comprehensively, and systematically.

\subsection{Magnetic star-planet interactions}
\label{sec:intro:mspi}
Magnetic SPI can occur if the planet is orbiting within the Alfv\'en radius of the star at least part of the time~\citep{preusse2006magnetic, cohen2011dynamics}. The Alfv\'en radius is the radius at which the stellar wind velocity, which increases with increasing distance from the star, exceeds the Alfv\'en velocity of the magnetized plasma. Beyond this radius, this plasma is disconnected from the star. Alfv\'en waves set off by a planet that crosses the magnetic field cannot propagate back to the star if the wind carries the plasma away faster than these waves can travel. In the sub-Alfv\'enic regime, the wave can reach the star, and deposit its energy in its atmosphere. \citet{lanza2012starplanet, lanza2018closeby}, and \citet{zarka2007plasma} and \citet{saur2013magnetic} suggest different mechanisms for when and how the energy transport and dissipation takes place, and how much energy can be transferred and dissipated at all. They all lack observational constraints from large samples for calibration, which we aim to provide in this work.

\subsection{Searching for global changes in activity}
\label{sec:intro:global}
Even if the instrument's sensitivity is sufficiently high, measuring magnetic star planet-interaction remains notoriously difficult. In individual systems, a non-detection may not mean an absence of the interaction, but merely a temporary cessation~\citep{shkolnik2005hot, shkolnik2008nature}. Statistical studies searching for changes in global chromospheric and coronal activity indicators in star-planet systems with close-in planets relative to planet-less stars are inconclusive~\citep{kashyap2008extrasolar,scharf2010possible, shkolnik2013ultraviolet, france2018farultraviolet, viswanath2020statistical, krejcova2012evidence, miller2015comprehensive, poppenhaeger2010coronal}. In these studies, it is challenging both to define a control sample to measure the activity against, and to quantify how the way we detect planets introduces selection bias in the activity measure. It is, for instance, easier to detect a planet around a magnetically inactive star, because both radial velocity and transit detections will be less affected by stellar variability~\citep{poppenhaeger2011correlation}, but that does not imply that SPI quenches activity. Further, tidal interactions can induce a global increase in activity~\citep{ilic2022tidal}, which adds a physical ambiguity to the interpretation~(see also Section~\ref{sec:discussion:tidal}). 

\subsection{Flares}
\label{sec:intro:flares}
In this study, we look at stellar flares as another way of dissipating the energy stored in the magnetic field. Flares are strong and impulsive eruptions released by reconnection and subsequent relaxation of field lines in the corona~\citep{svestka1976solar,priest2002magnetic}. In contrast to solar flares, stellar flares often are extremely energetic, sometimes enhancing the stellar flux by orders of magnitude~\citep{maehara2012superflares, shibayama2013superflares, paudel2018k2} so that they can be detected in stars over 1kpc away~\citep{chang2015photometric}. But since they occur randomly in time, individual events can be hard to catch. 

In the 2010s, a wealth of flare observations from space missions like Kepler~\citep{borucki2010kepler} and K2~\citep{howell2014k2} have been cataloged~\citep{davenport2016kepler, paudel2018k2, ilin2021flares}. Since 2018, the Transiting Exoplanet Survey Satellite~\citep{ricker2015transiting} is rapidly growing the archive~\citep{gunther2020stellar}. The key advantage of both missions for this work is the nearly uninterrupted monitoring of star-planet systems that covers all phases of a close-in planet's orbit.

\subsection{Flaring star-planet interactions}
\label{sec:intro:fspi}

\begin{figure}[ht!]
\script{paper_illustrative_sketch_of_system_geometries.py}
    \begin{centering}
        \includegraphics[width=0.9\linewidth]{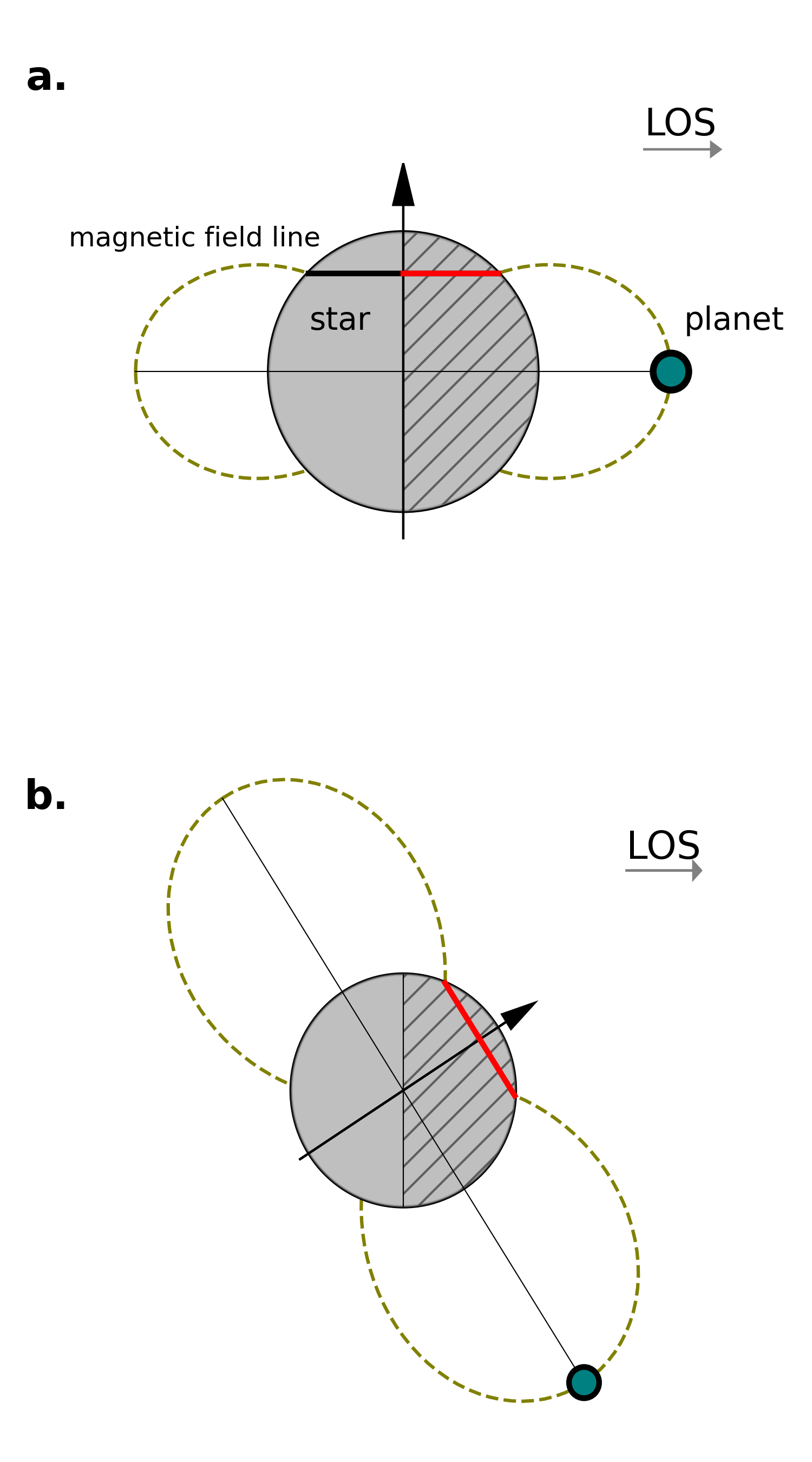}
        \caption{
         Simple magnetic star-planet interaction (SPI) viewing geometries. \textbf{a.: system aligned along the line of sight (LOS). } The interaction footpoint moves along the black line, and is on the visible hemisphere (hatched area) for $50\%$ of the  orbit (red line). Flaring SPI can be observed as modulation of flare occurrence times. \textbf{b.: Inclined system.} The interaction latitude is visible $100\%$ of the time, so no modulation can be observed. Interaction is shown on one hemisphere, but may appear on both. Complex geometries arise when orbital, rotational and magnetic axis are misaligned. The panels also illustrate how a planet in a closer orbit (a.) interacts at a lower latitude than a farther out planet (b.). }
        \label{fig:sketch}
    \end{centering}
\end{figure}

The absolute power of flaring SPI is difficult to constrain. We do not know how energetic planet-induced flares can be. The estimates for dissipated energy~\citep{lanza2018closeby} are in the ballpark of the regime of superflares, with energies released at optical wavelengths $\geq10^{33}\,$erg~\citep{schaefer2000superflares}, that can be observed by Kepler and TESS. However, on the one hand, the planet might be triggering flares prematurely, which would be of lower energy~\citep{loyd2023flares}, and end up below our detection threshold. On the other hand, the energy need not be dissipated continuously, particularly in the stretch-and-break mechanism proposed by~\citep{lanza2012starplanet}, so an individual flare's energy might be higher. We also do not know whether flares triggered by interactions with a companion would unfold faster or slower than intrinsic flares. We will assume that, since they would be caused by magnetic reconnection triggered by the planet in the stellar corona, they should still have the same physics as intrinsic flares. They should then approximalety follow the fast-rise-exponential-decay morphology~(see Fig.~\ref{fig:flares_and_fps} for an example). Individual intrinsic flares can deviate from this template, yet still be classified as flares because flares remain the best explanation for a sudden rise and drop in stellar brightness in most main sequence dwarf stars. Many of these deviations are unexplained today~\citep{howard2022no}, so that identifying SPI flares based on morphology is not possible without additional constraints. However, if we can statistically identify enough SPI flares, as we attempt in this work, we can infer the differences between intrinsic and SPI events from the two ensembles.

In other words, since planet-induced flares might not look any different from intrinsic flares, and we do not know how their energies will be distributed, we require a method to distinguish them without referring to their individual properties. Instead, we can look at the statistical distribution of flares in time: 

If we picture the magnetic field lines connecting the planet to the star in the sub-Alfv\'enic zone, we expect that the stellar footpoint of these lines will move along with the planet. Flares triggered by the planet will occur at a preferred orbital phase, namely whenever the footpoint is visible to the observer. Depending on the system's viewing geometry, i.e., the inclination of the rotation axis, the spin-orbit (mis-)alignment, and the extent and configuration of the large scale magnetic field -- the footpoint could be always in view, or never. In expectation, however, in star-planet systems with favorable viewing geometries and magnetic field strengths and configurations, the footpoint visibility will be modulated in phase with the planetary orbit~(Fig.~\ref{fig:sketch}). The resulting phase preference of planet-induced flares introduces a deviation from the otherwise random distribution of flares~\citep[see, e.g.,][]{doyle2018investigating,feinstein2020flare,howard2021evryflare}. Space missions like Kepler and TESS provide light curves that cover all phases of the planet's orbit, so that phase-correlated planet-induced flares can be detected against a background of uncorrelated intrinsic flaring.

\subsection{Searching for local changes in activity}
\label{sec:intro:local}
In contrast to global SPI, local SPI effects are easier to tell apart from intrinsic stellar behavior because they are tied to the planet's orbital period. Searches for an orbital modulation of polarized radio emission have produced promising, but tentative results. Close-in star-planet systems are expected to interact similarly to the Jupiter-Io system~\citep{bigg1964influence}, albeit scaled up to the high field strengths of stars~\citep{turnpenney2018exoplanetinduced}. In some detections, the planet has not been found (yet)~\citep{vedantham2020coherent}, or the data do not cover multiple orbital periods to see the signal repeat~\citep{perez-torres2021monitoring, pineda2023coherent}, or both \citep{callingham2021population}. Phase-correlated activity, flaring and otherwise, has been searched for in individual systems. After eight years of monitoring of HD 189733, an active Hot Jupiter host, higher X-ray energy flares seem only marginally clustered in orbital phase than lower energy ones~\citep{pillitteri2022xray}. \citet{maggio2015coordinated} observed an eruption of X-ray emission when the Jupiter-sized companion was in periastron on the highly eccentric orbit around its host HD 17156. However, a similar study using chromospheric indicators in the eccentric system HD 80606 found no variation between periastron and apoastron~\citep{figueira2016activity}. In the above cases, non-detection can at least partly be attributed to poor orbital phase coverage, which also limits the significance of any positive detection. Realizing this, \citet{fischer2019timevariable} used K2 data with excellent phase coverage of the orbits of the inner \mbox{TRAPPIST-1} planets to look for clustered flaring in phase, but found none. \citet{ilin2022searching} and \citet{klein2022one} found hints of excess flaring, and chromospheric Helium emission in the young AU Mic system, respectively. This mixed picture of individual object studies calls for a more comprehensive and systematic approach.

\subsection{Overview}
In this work, we searched for excess flaring in all available Kepler and TESS short cadence light curves of confirmed star-planet systems~(Section~\ref{sec:data}). We scoured the light curves for flares, and tested their observed phases for departures from a uniform distribution in phase with the innermost planet's orbit~(Section~\ref{sec:methods}). We present the catalog of all flares found in our sample of star-planet systems, and compare the test results with the expected power of interaction for each system~(Section~\ref{sec:results}). We discuss the results for the most interesting star-planet systems individually~(Section~\ref{sec:results:individualstars}), and take a look at the patterns emerging from our analysis~(Section~\ref{sec:discussion}), as well as the possible role of tidal flaring SPI in our observations~(Section~\ref{sec:discussion:tidal}).  We summarize our findings in Section~\ref{sec:summary}.

\section{Data}
\label{sec:data}
In this study, we aimed at compiling the largest possible sample to search for flaring SPI. We took the entire Planetary Systems Composite Parameters Table (PSCP\footnote{ \url{https://exoplanetarchive.ipac.caltech.edu/cgi-bin/TblView/nph-tblView?app=ExoTbls&config=PSCompPars}}, Section~\ref{sec:data:sps}) from the NASA Exoplanet Archive, as of July 2022, and scanned the systems for available Kepler and TESS light curves~(Section~\ref{sec:data:photometry}). We searched a total of 1811 systems and over 7200 light curves from both missions for flares. To calculate the clustering of flares in orbital phase, we further required the orbital periods of each system~(Section~\ref{sec:data:orbitalperiod}). We then estimated the theoretically expected power of interaction for each system with at least three flares. For this, we required further system parameters -- distance between planet and star, stellar rotation period, planetary radius, and stellar luminosity~(Sections~\ref{sec:data:a}-\ref{sec:data:lum}), which we mostly took from the literature. We began by compiling the table of system parameters with the PSCP, but verified, supplemented and updated the table with more up-to-date measurements. The relevant properties of the final sample of 25 systems are shown in Table~\ref{tab:maintable_lit}.

\begin{table*}
\begin{rotatetable*}
\script{paper_main_table.py}
 \caption{Flaring single star-planet system parameters. Obs. time is the total observing time with Kepler and TESS. The digits in square brackets refer to the uncertainties in the corresponding last digits shown, e.g., $0.051[1]$ is the same as $0.051\pm0.001$, and $98[12]$ is the same as $98\pm12$.}
\movetableright=-40mm
   
\footnotesize

    \input{output/table_lit_vals.tex}
        \label{tab:maintable_lit}
    \tablerefs{\input{output/lit_table_bibstring.tex}}
\end{rotatetable*}
\end{table*}

\subsection{Star-planet systems}
\label{sec:data:sps}
We compiled our sample of star-planet systems from the PSCP, from which we removed controversial planet detections (``pl\_controv\_flag'' must be 0), i.e., detections with existing literature questioning the result. The resulting catalog contained 2993 transiting and 191 non-transiting systems, from which we picked the innermost known planet in each system. 

\subsection{Kepler and TESS photometry}
\label{sec:data:photometry}
The Kepler and TESS missions are unbeaten with respect to long-term optical monitoring of stellar flares. Their excellent coverage of orbital phases makes the light curves ideal for our search for orbital phase dependent flaring. Between 2009 and 2013, the Kepler space telescope~\citep{koch2010kepler} nearly continuously observed a patch of the sky in the Cygnus-Lyra region. Each of the 18 observing Quarters contains nearly uninterrupted $\sim 90$ days of observations, totaling over $100\,000$ stars monitored in 2-min cadence in a broad $400-850\,$nm band pass.

The Transiting Exoplanet Survey Satellite~\citep[TESS,][]{ricker2015transiting} is an all-sky mission that began operations in 2018, and has completed two sky scans in summer 2022. It is observing at the time of writing, collecting nearly continuous photometric time series in the $600-1000\,$nm band for $\sim 27\,$d in each observing Sector. About $200\,000$ stars have been observed in 2-min cadence in the first two years of operations alone, with about $20,000$ targets per Sector. Out of these, from Sector 27 on, $1,000$ targets were observed at even higher 20-s cadence in each Sector. 

Based on the filtered PSCP table, we used the \texttt{lightkurve}~\citep{lightkurvecollaboration2018lightkurve} Python software to query the full Kepler archive (quarters Q0-Q17, Data Release 25), and the most recent TESS catalog (July 2022) for their respective 1-min and 2-min/20-s cadence light curves. In total, we searched 7213 light curves for flares -- 3032 Kepler Quarters, and 4181 TESS Sectors. We found and analyzed light curves for a total of 1811 systems, 344 of which were observed only by the primary Kepler mission, 1205 only by TESS, and 262 by both missions. We did not use K2 light curves because of the many systematics, and consequently the high time investment of de-trending them. But we included the flare list obtained by~\citet{paudel2018k2} for TRAPPIST-1 K2 flares, since this system is a potential candidate for sub-Alfv\'enic interactions~\citep{fischer2019timevariable}.

\subsection{Orbital periods and transit mid-times}
\label{sec:data:orbitalperiod}
We adopt the orbital periods from the PSCP, which were either obtained from Kepler or TESS transits, or from radial velocity measurements. For the transiting systems, we use the transit mid-times given in the PSCP to set orbital phase zero. If possible, we use the transit mid-times determined in each mission separately to reduce the uncertainties on orbital phase. For the non-transiting system, phase zero is set arbitrarily.

\subsection{Semi-major axes and eccentricities}
\label{sec:data:a}
A critical parameter for the possibility of magnetic SPI and its power is the distance between the star and the planet, not the semi-major axis itself. This varies if the orbit is eccentric. We therefore adopt the mean semi-major axes from the literature, but use a custom estimate for the uncertainty that includes the eccentricity:

If the eccentricity is known, then the uncertainty on the distance is set to either the error on the semi-major axis, or to half of the difference between periastron and apoastron -- whichever is larger. If eccentricity is not known, then the uncertainty on the distance is set to either a 25\% error on the semi-major axis, that is, assuming $e=0.25$, or the uncertainty on the semi-major axis -- whichever is larger. We chose $e=0.25$ because it is both a typical value within our sample~(see Table~\ref{tab:maintable_lit}), and in the literature~\citep{eylen2019orbital}.

\subsection{Rotation periods}
\label{sec:data:rotationperiods}
We adopt available rotation period values from the literature. Almost all stem from light curve variability~\citep{angus2018inferring, mazeh2015photometric, mcquillan2013stellar, mcquillan2014rotation, luger2017sevenplanet, stock2020carmenes, deleon202137, torres2017validation, stefansson2020habitable, zicher2022one, ment2021toi, rizzuto2020tess, gunther2020stellar}, and only one from periodic variation in chromospheric lines~\citep{demangeon2021warm}. When uncertainties are not given, we conservatively assume $10\%$ uncertainty. Uncertainties are missing usually when periods were detected using Lomb-Scargle~\citep{lomb1976leastsquares, scargle1982studies}, or similar, periodogram peaks only~\citep{gunther2020stellar, kiraga2007agerotationactivity, grankin2013magnetically, burt2014lickcarnegie}, but also when the measurement is indirect using activity-rotation relations~(only in the case of GJ 3323).

Among the stars without given uncertainties, the rotation of GJ~3082 was measured and is consistent both in TESS and in KELT light curves to below 1\,s precision~\citep{gunther2020stellar}. The rotation of GJ~674 is consistent between light curve periodograms~\citep{kiraga2007agerotationactivity}, and activity-rotation relations~\citep{boisse2011disentangling} at a $<10\%$ level. GJ~3323 only has rotation periods inferred from activity indicators~\citep{astudillo-defru2017magnetic}, without any periodicity detected in the activity indicators themselves~\citep{astudillo-defru2017harps}. The rotation of Proxima~Cen is very slow, but consistent between different datasets~\citep{anglada-escude2016terrestrial, kiraga2007agerotationactivity}. In general, among the stars without given uncertainties on the rotation period, the period is either 
\begin{itemize}
    \item short, leaving a clear peak in the periodogram, and little doubt about the period after ruling out aliases with $P_{\rm rot}/2$, as with GJ~3082, or
    \item longer than the Sun's, that is, in the regime where rotation-activity relations are reliable because the stars are in the unsaturated activity regime, as in the case of GJ~674, and GJ~3323.
\end{itemize} 


\subsection{Planetary radii}
\label{sec:data:planetradii}
For the transiting planets, we adopt the literature values for planet radius $R_{\rm p}$, and uncertainties, from the PSCP. For the non-transiting radial velocity detected planets we use  $M_{\rm p}\sin i$ to calculate a lower limit for the radius using the empirical relations derived in \cite{chen2017probabilistic} using their open source \texttt{forecaster} tool, upgraded to \texttt{astro-forecaster}\footnote{https://pypi.org/project/astro-forecaster/} by Ben Cassesse. We note that for GJ 674, the mass estimates in \cite{bonfils2007harps} and \cite{boisse2011disentangling} do not quote uncertainties, but differ by $0.3M_\oplus$, so we assumed that value as the uncertainty on $M_{\rm p}\sin i$.

\subsection{Bolometric luminosity}
\label{sec:data:lum}

We take bolometric luminosity values as given in the PSCP whenever they are given with uncertainties. We supplement missing values, and replace entries without quoted uncertainties with Gaia DR3 FLAME~\citep{fouesneau2022gaia} solutions~(\texttt{lum\_flame, lum\_flame\_upper, lum\_flame\_lower}). 

\section{Methods}
\label{sec:methods}
We measure flaring SPI as the presence of flares triggered by the orbiting planet. In the absence of flaring SPI, flare peak times will be distributed randomly in orbital phase. In the presence of flaring SPI, we measure a phase dependent deviation from this randomness.

The main data for this analysis are flare times. To obtain them, we gather the Kepler and TESS light curves for all star-planet systems, remove rotational variability trends, search the de-trended light curves for flares, and convert the flare times to orbital phases. We then perform a customized Anderson-Darling test on the flare peak time distribution, which yields a $p$-value for the significance of the SPI signal. Finally, we compare these results to the theoretically expected SPI power $P_{\rm spi}$ in each system.

The methods for light curve de-trending and flare finding in Kepler and TESS light curves, as well as the Anderson-Darling test, are the same as detailed in~\citet{ilin2022searching}. We briefly recap the techniques in Sections~\ref{sec:methods:flaresearch} and \ref{sec:methods:adtest}. The expected power of SPI depends on the relative velocity between the planet and the magnetic field strength in its orbit, the derivation of which we explain in 
Sections~\ref{sec:methods:relvel} and \ref{sec:methods:bfield}, respectively. We can then combine them with the stellar radius, planetary radius, and semi-major axis to estimate the power of SPI. We use the scaling laws for two different magnetic SPI models, which we introduce in Section~\ref{sec:methods:pspi}.

\subsection{Light curve de-trending and flare search}
\label{sec:methods:flaresearch}
To remove trends and rotational variability from the light curve without losing the flare signal, we use an empirically derived multistep process, implemented as the \texttt{custom\_detrending} method in \texttt{AltaiPony}, a flare science Python package for light curve analysis~\citep{ilin2021altaipony}. First, we apply a spline fit with a coarse sampling of 30h-averaged values to capture slow rotation with periods above multiple days and non-periodic trends. Then, we iteratively fit a series of sines to capture rotational signal on time scales down to about half a day, close to the fastest rotational signals measured in low mass stars. Eventually, we apply two Savitzky-Golay~\citep{savitzky1964smoothing} filters in sequence, with window sizes of 6h and 3h each. At this stage, we mask all data points above a $2.5 \sigma$ threshold (or $1.5 \sigma$ for stars like AU Mic or Proxima Cen, which are very active or have very low noise levels, or both) as flare candidates to prevent the filter from ironing out the flares. As a final step, we fit exponential functions to the edges of the light curves, if there are data points that deviate more than one standard deviation from the median value, while keeping the flare candidates masked. Transits are usually too shallow to affect either light curve de-trending or flare finding. The light curve portions around deep transits were inspected manually.

In the de-trended light curves, we search for flares as series of at least three consecutive data points $3\sigma$ above the noise level using the \texttt{AltaiPony} method \texttt{find\_flares}. We estimate the noise level as the standard deviation in a rolling window of two hours, while masking deviations above $2.5\sigma$ (or $1.5\sigma$  for stars like AU Mic and Proxima Cen). To capture the exponential tail of the flare, we use the \texttt{addtail} flag in the \texttt{find\_flares} method to continue adding data points to the end of the flare as long as they are $2\sigma$ above the noise threshold.

We vet all flare candidates by eye, and exclude instrumental and physical false positives, such as Solar System Objects passing by the field of view~(see Fig.~\ref{fig:flares_and_fps} in the Appendix for an example). For each confirmed flare candidate, we calculate the relative amplitude $a$ and equivalent duration $ED$, defined as the flare flux $F_{\rm flare}$ over the duration of the flare, divided by the median flux $F_0$ of the star, integrated over the flare duration~\citep{gershberg1972results}:
\begin{equation}
\label{eq:ED}
ED=\displaystyle \int \mathrm dt\, \frac{F_{\rm flare}(t)}{F_0}.
\end{equation}
In other words, equivalent duration is the time during which the non-flaring star releases as much energy as the flare.

\subsection{Custom Anderson-Darling test}
\label{sec:methods:adtest}
For each star-planet system with three or more flares in its light curves, we test for deviations from a random distribution of flares with orbital phase. We use the same customized Anderson-Darling test as in \cite{ilin2022searching}, which is more sensitive to deviations at both ends of the distribution than the widely used Kolmogorov-Smirnov test~\citep{feigelson2012modern}. In short, we first take the number of flares observed in the TESS and Kepler data to calculate a base flare rate for each light curve. We then calculate how often each orbital phase has been observed, also for each light curve separately, to ensure that the different detection thresholds for TESS and Kepler do not bias the base rate. With the phase coverage and base flare rate combined, we can tell how many flares we would expect to see in any given phase bin if flares were randomly distributed. The bin widths correspond to the distance between each consecutive detection of a flare in phase space. We then aggregate this number of expected flares per orbital phase bin into one expected distribution (blue curves in Figs.~\ref{fig:cumdist_transiting} and \ref{fig:cumdist_rv}) by summing over all available light curves. As a last step, we compare the expected distribution to the observed one using an Anderson-Darling test. For this, we sample 10000 times from the expected distribution the same number of flares as we observe in each system. For instance, for HIP 67522, we sample 12 flares distributed according to its blue curve in Fig.~\ref{fig:cumdist_transiting}. For each of the 10000 samples, we compute the Anderson-Darling statistic $A^2$, which yields a distribution of $A^2$ values~(see~Fig.~\ref{fig:ad_dist_hip} for the $A^2$ distribution derived for HIP~67522). In a last step, we compute $A^2$ for the observed distribution, and compute the $p$-value. We repeat this test with different phase offsets to account for remaining biases in sensitivity of the test at different phases.

The only adjustment to the procedure in \cite{ilin2022searching} is that we use four equidistant start phases (i.e., 0, 0.25, 0.5 and 0.75) compared to the 20 used in~\cite{ilin2022searching} to save time, because the range of derived $p$-values is already well-sampled using four. We adopt the standard deviation of these four $p$-values as the uncertainty on the flaring SPI measurement.

\subsection{Relative velocity}
\label{sec:methods:relvel}
We calculate the relative velocity between the magnetic field of the star and the planet using stellar rotation period $P_{\rm rot}$, orbital period $P_{\rm orb}$, and semi-major axis $a$. We assume that the large scale magnetic field is co-rotating with the stellar surface, and calculate the relative velocity at the planet's orbital distance:

\begin{equation}
    v_{\rm rel} = 2 \pi a \left(\frac{1}{P_{\rm orb}} - \frac{1}{P_{\rm rot}}\right).
\end{equation}

We use quadratic error propagation to estimate the uncertainty in $v_{\rm rel}$ with the mean of the upper and lower uncertainty values on the orbital period and rotation period, if both are given, and the uncertainty on semi-major axis $a$ as derived in Sec.~\ref{sec:data:a}. 

\subsection{Stellar magnetic fields}
\label{sec:methods:bfield}
We derive the average magnetic field strength $B$ from Rossby number $R$o using the empirical relation derived in \citet{reiners2022magnetism},  Table 2, in the unsaturated and saturated regimes, respectively:

\begin{eqnarray}
    B &= 199\,\text{G} \cdot R\mathrm{o}^{-1.26\pm 0.1} \;(\text{if}\; R\mathrm{o} > 0.13) \label{Eq3}\\
    B &= 2050\,\text{G} \cdot R\mathrm{o}^{-0.11\pm 0.03} \;(\text{if}\; R\mathrm{o} < 0.13) \label{Eq4}
\end{eqnarray} 

The convective turnover time $\tau$ in  $R \mathrm{o}=P_{\rm rot}/\tau$ is derived using bolometric luminosity~(Section~\ref{sec:data:lum}), following~\citet{reiners2014generalized, reiners2022magnetism}:

\begin{equation}
    \tau = 12.3\, \mathrm{d} \cdot (L_{\rm bol}/L_\odot)^{-1/2}
\end{equation}

\begin{figure}[ht!]
    \script{paper_lx_plots.py}
    \begin{centering}
        \includegraphics[width=\linewidth]{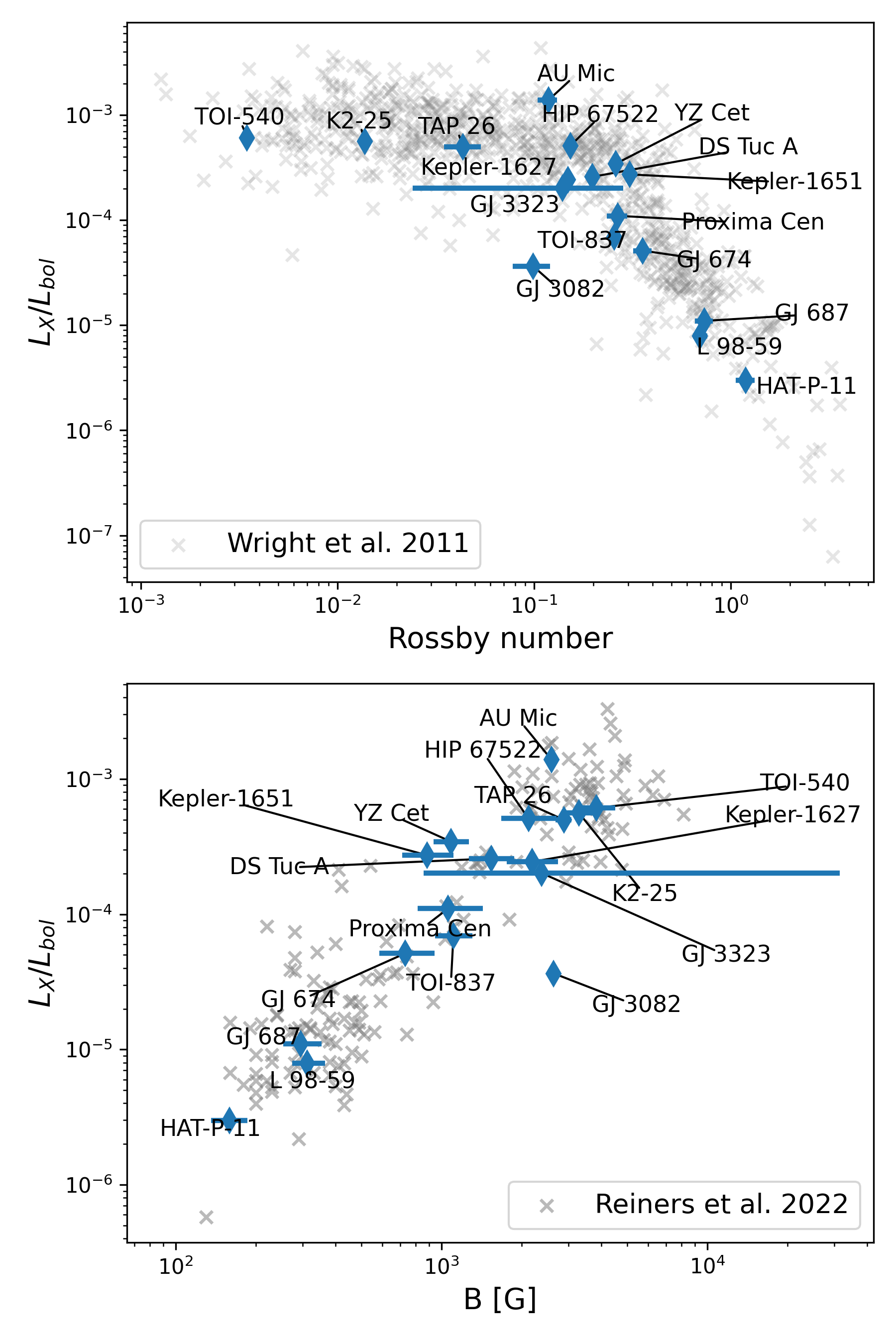}
        \caption{
           X-ray luminosity over bolometric luminosity compared to Rossby number (\textbf{upper panel}), and average surface magnetic field strength (\textbf{lower panel}). Blue lines denote the error bars, black lines point to the IDs of the corresponding stars. Both relations follow those of stars not selected for hosting close-in planets, except for GJ 3082, which is underluminous for its Rossby number, and hence also its estimated magnetic field appears too strong for its $L_X/L_{\rm bol}$.
        }
        \label{fig:lx}
    \end{centering}
\end{figure}

We compared our derived values for $R$o and $B$ with existing estimates of coronal emission. We show $L_X/L_{\rm bol}$ obtained from~\citet{foster2022exoplanet} as a function of $R$o and $B$ in Fig.~\ref{fig:lx}. The X-ray emission relation to Rossby number in our sample follows that of stars that are not known to host close-in planets~\citep{wright2011stellaractivityrotation}, spanning both the saturated and unsaturated regime. The X-ray emission and our estimated magnetic field closely follow each other, also consistent with stars not selected for close-in planets~\citep{reiners2022magnetism}. There are two exceptions: GJ 3082 appears underluminous in X-ray compared to its Rossby number and expected magnetic field strength. AU Mic is slightly overluminous, falling somewhat above the saturated level of $L_X/L_{\rm bol}$. However, the star still follows the $B-L_X/L_{\rm bol}$ relation.  

\subsection{Power of star-planet interaction}
\label{sec:methods:pspi}
We consider two theories for the mechanism behind flaring SPI, the stretch-and-break and the Alfv\'en wing mechanisms. For both of them, we include the case of a magnetized and an unmagnetized planet, for a total of four estimates of the expected power of magnetic SPI.

First, we estimate the power of SPI generated by the stretch-and-break mechanism using the scaling relations in~\cite{lanza2012starplanet}. We adopt their Eqn.~45 for non-linear and axisymmetric magnetic fields, because the non-linear formulation does not assume a constant force-free parameter, and because only axisymmetric fields can produce modulation of flaring activity in phase with the planetary orbit. We note that the axisymmetry of stellar magnetic fields varies across stars~\citep[e.g.,][]{donati2009magnetic, donati2008largescale, morin2008largescale, morin2010largescale}, and also with the stellar activity cycle of individual stars~\citep[e.g.,][]{borosaikia2016solarlike, lehmann2021identifying}. In Eqn. 45, the power of SPI, $P_{\rm spi,sb}$, is

\begin{equation}
      P_{\rm spi,sb}=\underbrace{\frac{27\pi \lambda^2(n+1)}{16 \mu_0 n (\lambda^2 + n^2)^{1/3}}}_{\hat{C}} B_{\rm p}^{2/3} R_{\rm p}^2 B_*^{4/3}  v_{\rm rel} \left(\frac{R_*}{a}\right)^{\frac{n+11}{ 3}},
\end{equation}

where $\mu_0$ is the magnetic permeability of the vacuum, and $n$ and $\lambda$ are dimensionless constants, all of which we subsume under $\hat{C}$, together with other constant factors. $B_{\rm p}$ and $B_*$ are the planetary and stellar field strengths at their poles, respectively; $v_{\rm rel}$ is the relative velocity between the stellar rotation and the planet's orbit at the semi-major axis $a$; and $R_*$ and $R_{\rm p}$ are the stellar and planetary radii, respectively. We choose $n=0.25$ ($\lambda^2 = 1.01203$), which best reproduces observations of SPI candidate systems HD~179949, $\tau$ Boo, and HD~189733 in~\citet{lanza2012starplanet} (except for $\upsilon$ And). Since the planetary fields are unknown, we adopt $B_{\rm p}=1\,$G for simplicity. We also do not know the stellar field strength at the pole for the vast majority of the stars, so we adopt an average value of 15 per cent of the surface field strength~(see Sections~\ref{sec:methods:bfield} and~\ref{sec:discussion:as}) as a proxy.

Second, we also consider the case of an unmagnetized planet with $B_{\rm p}=0$, in which case Eq.~45 in~\cite{lanza2012starplanet} reduces to

\begin{equation}
      P_{\rm spi,sb0} =\hat{C} R_{\rm p}^2 B_*^2 v_{\rm rel}  \left(\frac{R_*}{a}\right)^{\frac{n+11}{ 3}}
\end{equation}

Third, we estimate the power of SPI generated by the Alfv\'en wing mechanism ~\citep{saur2013magnetic,kavanagh2022radio},  combining Eqn.~8 for the power of interaction and Eqn.~11 for the magnetopause radius of the planet from~\citet{kavanagh2022radio} to arrive at:

\begin{equation}
    P_{\rm spi,aw} = \frac{\pi^{1/2}}{2^{2/3}} B_{\rm p}^{2/3} R_{p}^2  R_*^2  B_*^{1/3}  v_{\rm rel}^2 a^{-2} \rho_*^{1/2} \sin^2 \theta
\end{equation}

Here, we assume that the magnetopause radius extends to where the magnetic pressures of planet and stellar magnetic wind are equal, and that the wind is dominated by the magnetic field in the sub-Alf\'enic regime. We approximate the stellar wind density $\rho_W$ with a proportionality to $\rho_* (R_* / a)^{2}$, where $\rho_*$ is the stellar wind density at the base of the corona. For the stellar wind magnetic field at the planetary radius $B_W$, we insert $B_* (R_* / a)^{3}$, i.e., we assume that the field strength decreases radially outward from the surface as a dipole. This produces an upper bound on the field strength at the planet's distance because more complex fields would drop off more steeply. The last two replacements result in the total $R_*^2 a^{-2}$ factor in the above equation. Since $\rho_*$ is poorly constrained empirically~\citep{vidotto2021evolution}, we adopt $\rho_*=2\cdot10^{10}\rm{cm}^{-3}$, a value often used for the Sun~\citep{sokolov_magnetohydrodynamic_2013}, and for stellar simulations~\citep{alvarado-gomez2020tuning,kavanagh2021planetinduced}. The dependence on $\rho_*$ is stronger than that on $B_*$, so it may dominate the magnetic field dependence. However, it is also weaker than that on any other parameter in the scaling law. The angle between the wind magnetic field vector and the velocity vector of the planet $\theta$ is also unknown, so we set $\sin\theta=1$ to obtain an upper limit. In reality, $\theta$ will vary both between systems, and over time in any individual system.

And fourth, we also consider the case of an unmagnetized planet,  where the magnetospheric radius is set equal to the planetary radius, in which case Eq.~8 in \cite{kavanagh2022radio} reduces to

\begin{equation}
    P_{\rm spi,aw0} = \pi^{1/2} R_{\rm p}^2 R_*^4 B_*  v_{\rm rel}^2 a^{-4} \rho_*^{1/2} \sin^2 \theta
\end{equation}


\section{Results}
\label{sec:results}

Our goal was to measure magnetic SPI as the statistical clustering of flares in phase with the innermost planet's orbit. We searched for flares in all star-planet systems that were observed with Kepler and TESS at 1 or 2 min cadence, respectively. For the star-planet systems in the resulting flare catalog~(Section~\ref{sec:results:catalog}), we confirmed that the orbital phase could be known well enough for the entire observing baseline of the system~(Section~\ref{sec:results:coherence}). We calculated the orbital phases of each flare, and the flare phase distributions for each system~(Section \ref{sec:results:phasedist}). We then applied a custom Anderson-Darling test~(see Section~\ref{sec:methods:adtest}) to assess how much each distribution was different from randomly distributed intrinsic flaring. Comparing the test results with the expected power of magnetic SPI, we found our main result: The clustering of flares in orbital phase tentatively increases with the expected power in all considered scenarios of magnetic SPI~(Section~\ref{sec:results:spi}). However, not all systems with high expected power show signs of flaring SPI, which creates two branches in the distribution of test results. Finally, in Section~\ref{sec:results:individualstars}, we provide additional context for the most interesting systems in our sample, and explain why we had to exclude others from the analysis.

\subsection{Flare catalog}
\label{sec:results:catalog}
We searched a total of 3032 Kepler Quarters and 4181 TESS Sectors of a total of 1811 star-planet systems for flares. We inspected all candidates manually, and added flares observed by K2 on TRAPPIST-1~\citep{paudel2018k2}, and flares from the planet hosting primary in the Kepler-411 binary~\citep{jackman2021stellara}. The final table contains a total of \input{output/PAPER_total_number_of_flares.txt}flares in \input{output/PAPER_total_number_of_systems_with_flares.txt}systems. In Table~\ref{tab:flares}, we list the flares and their characteristics. To identify each event, we provide the name of the system in which the flare was found, its TIC, whether it was observed with Kepler or TESS, and during what Quarter or Sector, respectively. For each flare, we also give the start and finish time, relative amplitude $a$ and equivalent duration $ED$~(Eq.~\ref{eq:ED}). If the transit midtime is known for the innermost planet, we also give the orbital phase at which the flare occurred. The full table is available online (see Data Availability Statement).

\begin{table*}
    \script{paper_latex_flare_table.py}
    \centering
            \caption{
            Flare catalog of all star-planet systems observed by Kepler and TESS (as of July 2022). In transiting multi-planet systems, the orbital phase refers to the innermost planet, with the transit mid-time at phase zero. Qua./Sec. is the Quarter or Sector number of the observation in Kepler or TESS, respectively. The first and last data points of a flare are denoted $t_s$ and $t_f$. The orbital phase of the innermost planet is measured at $t_s$. We also give relative amplitude $a$, and equivalent duration $ED$ (see Eq.~\ref{eq:ED}) for each flare. The digits in square brackets refer to the uncertainties in the corresponding last digits shown, e.g., $0.051[1]$ is the same as $0.051\pm0.001$, and $98[12]$ is the same as $98\pm12$. The full catalog is available online (see Data Availability Statement).
        }
    \input{output/flare_table.tex}
        \label{tab:flares}
\end{table*}

\subsection{Period coherence times}
\label{sec:results:coherence}
In many cases, the observing baseline covered by Kepler and TESS for a given system can span multiple years. In these cases, the orbital period of the innermost planet must be known very precisely so that we can assign accurate orbital phases to the flare events.
We test this by dividing the timespan $\Delta T$ between the first and the last flare in the combined Kepler and TESS observations by the coherence time $\tau$ of the orbital period for each system. We calculate the coherence time as

\begin{equation}
    \tau = P^2 / \sigma_P,
\end{equation}
where $P$ is the orbital period, and $\sigma_P$ is its uncertainty.
If the orbital period is well-known, the resulting ratio $\Delta T/\tau$ should be $\ll 1$. Here, we are not interested in the absolute orbital phase. We are only interested in how much uncertainty on the orbital phases of flares accumulates over the entire observing time after the first flare. Therefore, we can ignore that the uncertainty on phase zero at transit midtime increases with the time passed since the transit midtime was measured. For non-transiting systems, this time can amount to several years, so the midtime becomes unconstrained by the time of the observation with Kepler or TESS. For transiting systems, the midtime is usually very well known. So in the following, we set phsae zero to an arbitrary time in non-transiting systems, and to the reported transit midtime for the rest. 

\begin{figure}[ht!]
    \script{paper_coherence_histogram.py}
    \begin{centering}
        \includegraphics[width=\linewidth]{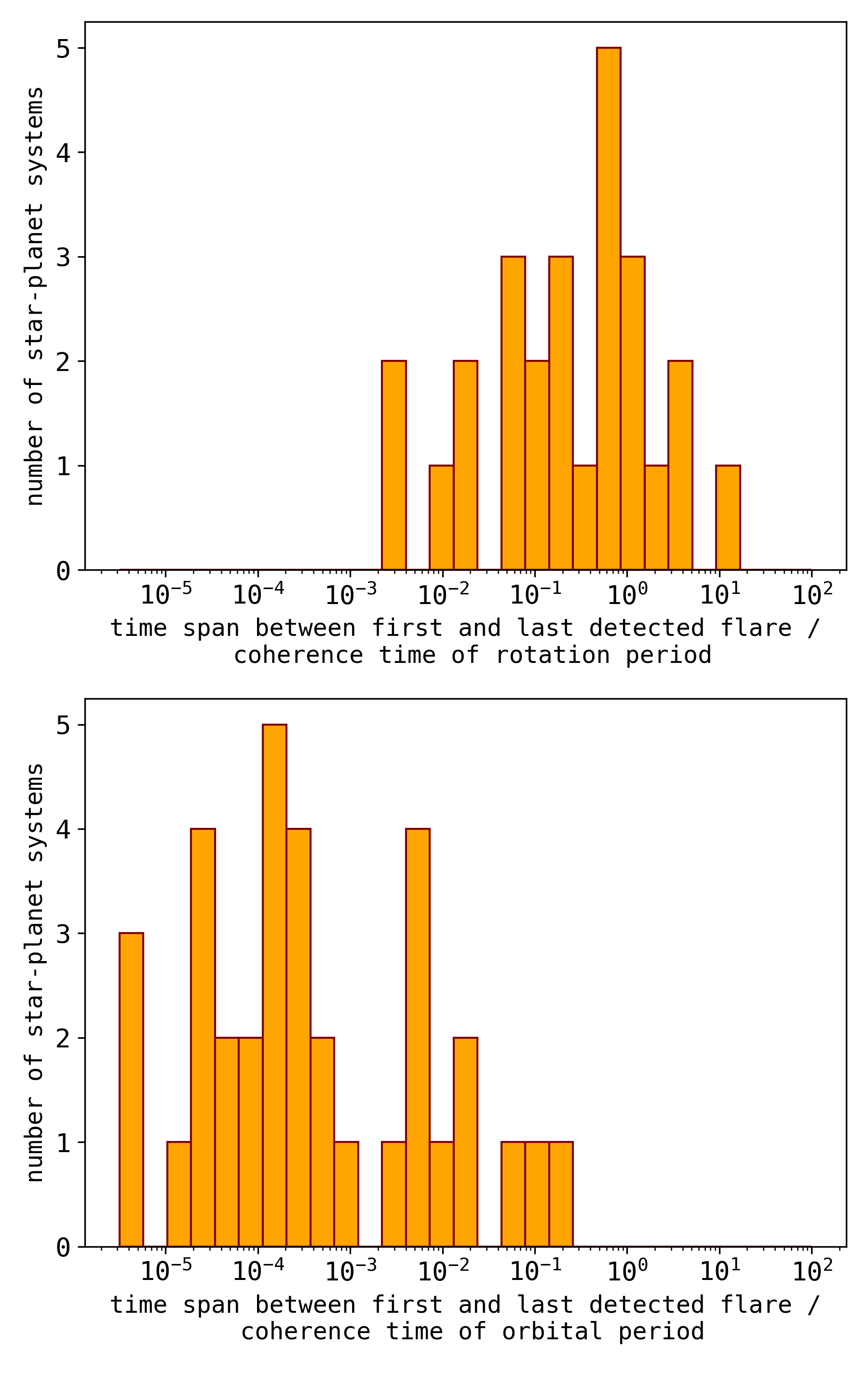}
        \caption{
           Time span of observation vs. coherence time of the rotational (\textbf{upper panel}) and orbital (\textbf{lower panel}) periods, respectively. Orbital periods are typically known precisely enough, so that the phase uncertainty at the last observed flare is of the order of $10^{-2}$. In contrast, rotation periods are usually less well constrained, so that the phase of the last flare is often undetermined (ratio on x-axis $\geq 1$).
        }
        \label{fig:coherence_hist}
    \end{centering}
\end{figure}

The bottom panel in Figure~\ref{fig:coherence_hist} shows a histogram of $\Delta T/\tau$ in our sample. The orbital period is known sufficiently well for our analysis in most cases with ratios $<0.02$, that is, the orbital phase of the last flare is at most $2\%$ off, compared to the first. The three cases where the ratio is highest with $\Delta T/\tau=$$0.05$, $0.12$, $0.24$ are TAP 26, GJ 3082, and GJ 674, respectively. We keep these systems in our sample, but caution that the measured absence of flaring SPI might be partly due to the uncertain orbital period of the planet in these cases.

\citet{fischer2019timevariable} argue that the synodic period, at which the sub-planetary point crosses the same (magnetic) surface element on the star, may result in a clearer interaction signal, because a magnetically active region that is prone to planet-induced flares, will be passed only once per that period. However, we repeated the coherence calculation for $P_{\rm rot}$, and found that the ratio is $>0.1$ for most stars~(Figure~\ref{fig:coherence_hist}, top panel), so that a statistical analysis of rotational flare modulation is not feasible. 

\subsection{Flare phase distributions}
\label{sec:results:phasedist}
\begin{figure*}[ht!]
    \script{paper_cumdist_individual_transiting.py}
    \begin{centering}
        \includegraphics[width=\linewidth]{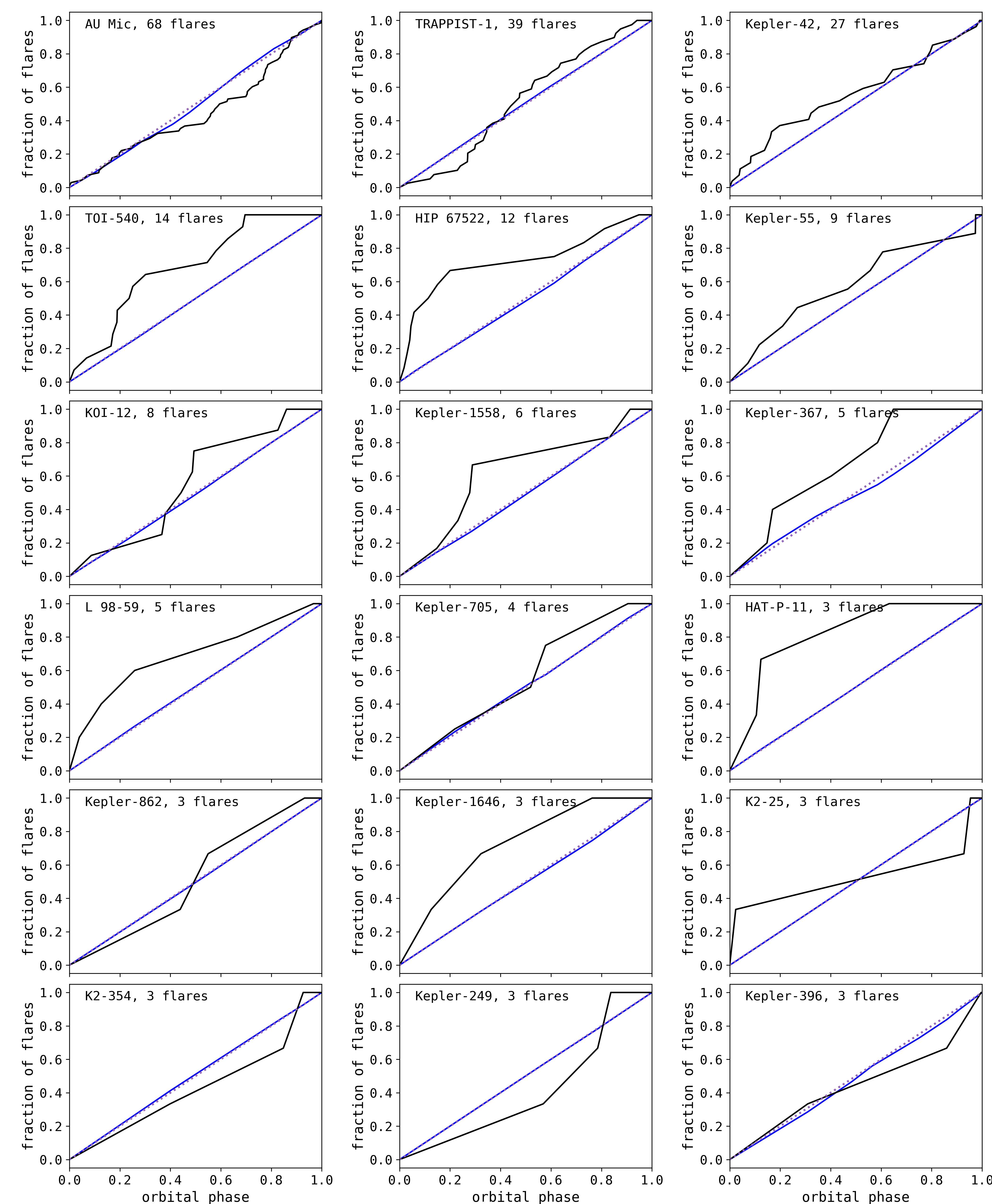}
        \caption{
            Cumulative distributions of orbital phases of flares in the \textit{transiting} planet hosts observed by Kepler and TESS, sorted by number of flares from top to bottom. The bisector line is dotted, the expected distribution is solid blue, and the observed distribution is solid black. Phase zero corresponds to the transit mid-time of the planet. 
        }
        \label{fig:cumdist_transiting}
    \end{centering}
\end{figure*}

\begin{figure*}[ht!]
    \script{paper_cumdist_individual_rv.py}
    \begin{centering}
        \includegraphics[width=\linewidth]{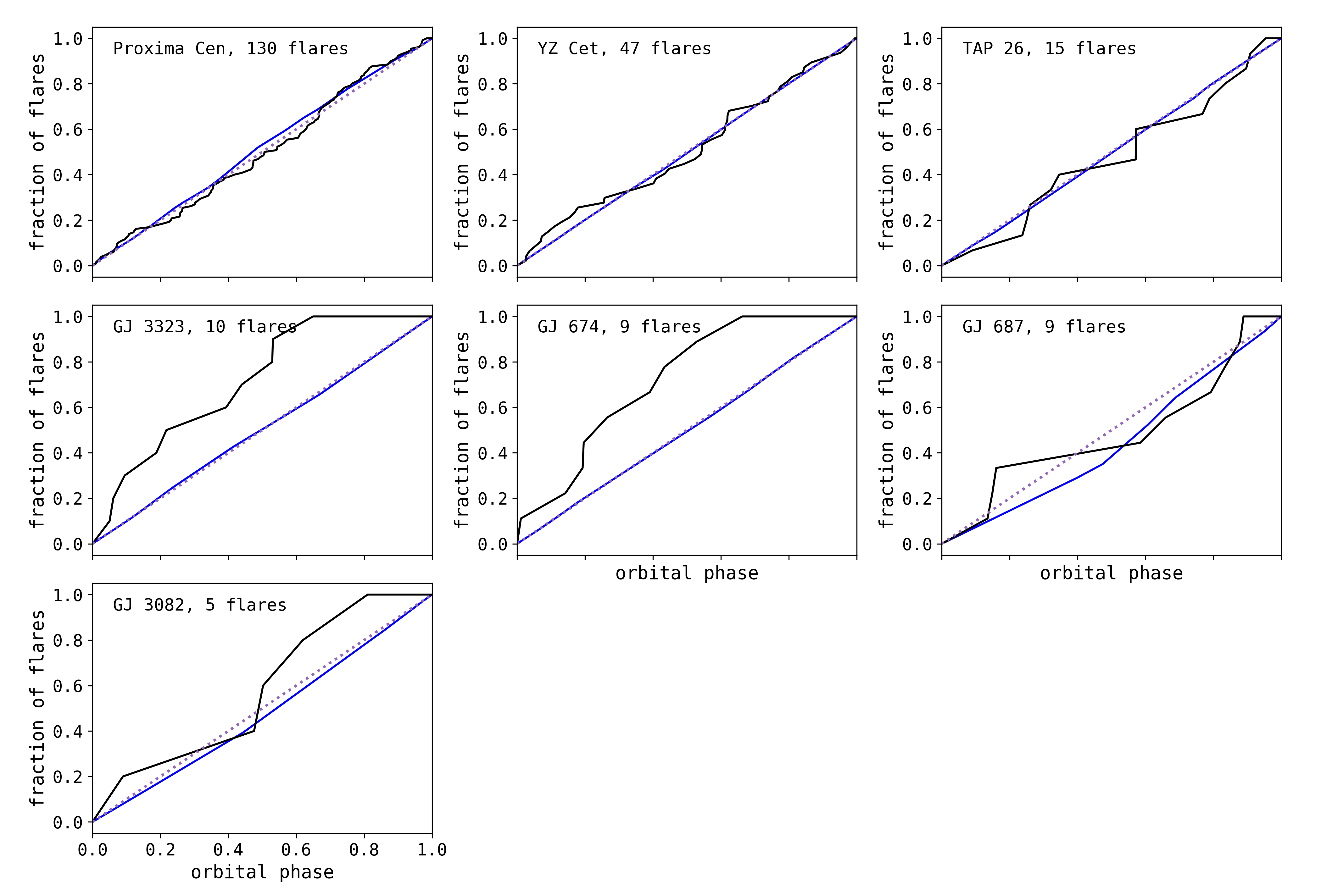}
        \caption{
            Cumulative distributions of orbital phases of flares in the \textit{non-transiting} planet hosts observed by Kepler and TESS, sorted by number of flares from top to bottom. The bisector line is dotted, the expected distribution is solid blue, and the observed distribution is solid black. Phase zero is chosen arbitrarily. 
        }
        \label{fig:cumdist_rv}
    \end{centering}
\end{figure*}

Our custom Anderson-Darling test compares the measured distribution of orbital phases of the flares to the expected distribution~(see Section~\ref{sec:methods:adtest}), and returns the significance of the deviation between the two. In the expected distribution, the same number of flares would be distributed randomly across all phases, as the overwhelming majority of studies looking for variation with rotational phase suggest~\citep[see, e.g.,][]{doyle2018investigating,howard2021evryflare}. Figures~\ref{fig:cumdist_transiting} and \ref{fig:cumdist_rv} show the flare phase distributions along with the corresponding expected distributions for transiting and non-transiting systems, respectively. The expected distributions usually deviate from a straight line because we take into account the coverage of the orbital phases by the Kepler and TESS observations, as well as the different flare rates in each light curve, which arise due to varying noise levels between individual Quarter/Sectors, and between the two instruments.

\subsection{Flaring star-planet interaction signal}
\label{sec:results:spi}
\begin{figure*}[ht!]
    \script{paper_adtest_vs_value_scatterplots.py}
    \begin{centering}
        \includegraphics[width=\linewidth]{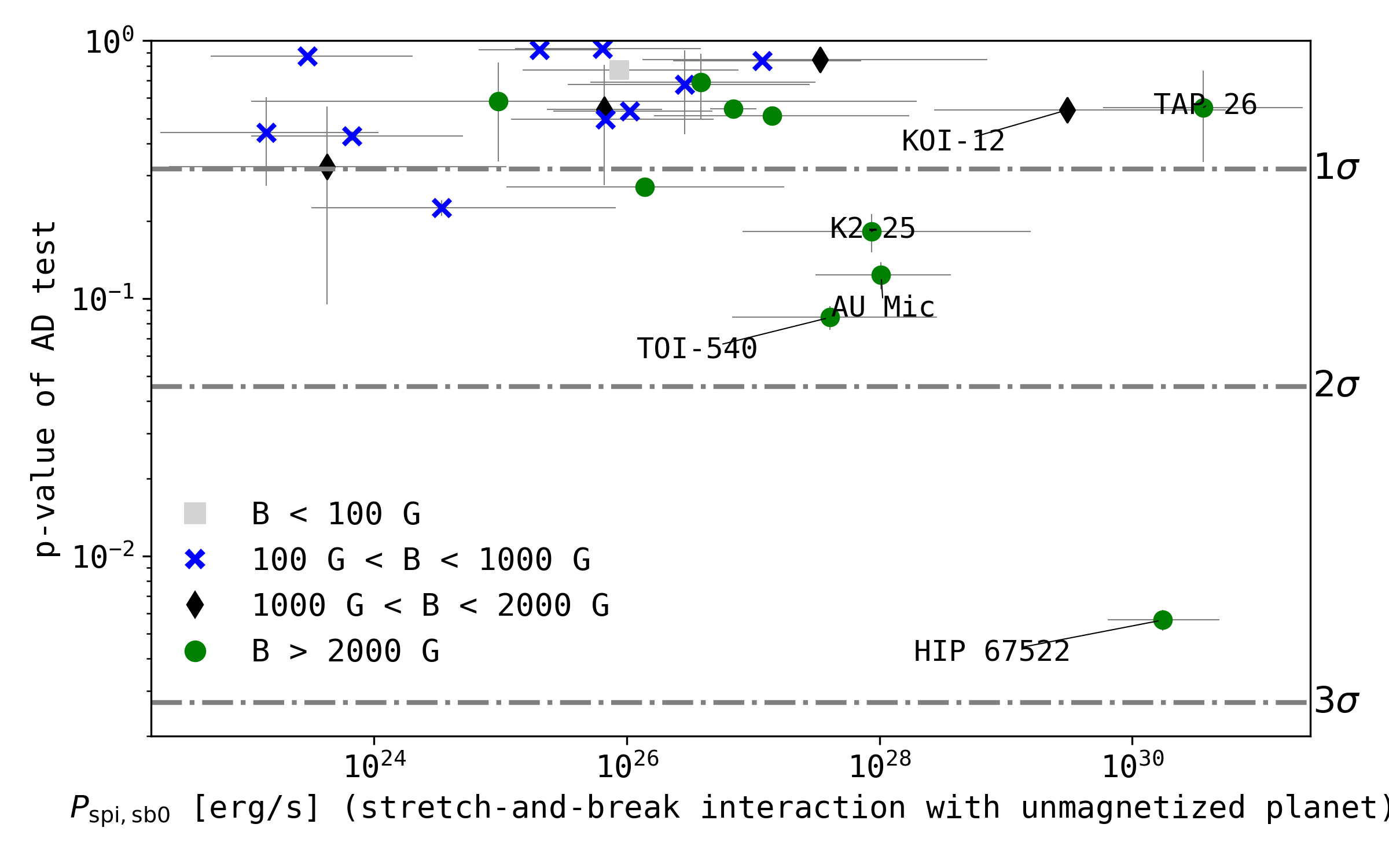}
        \caption{
            Expected power of magnetic SPI vs. AD test results, assuming the stretch-and-break scenario with an unmagnetized planet. The vertical and horizontal grey lines denote the uncertainties quoted in Table~\ref{tab:maintable_der}. While no individual system amounts to a $>3\sigma$ detection, the global trend is intriguing: Systems with low expected power show no deviation from random intrinsic flaring (i.e. high $p$-values in the left half of the Figure). In contrast, systems with high expected power show two branches. One where the significance of interaction increases with power, and one without such a trend. See Fig.~\ref{fig:adtest_ro} in the Appendix for the same figure, color-coded by Rossby number.
        }
        \label{fig:adtest_bp}
    \end{centering}
\end{figure*}

The main result of this work is the comparison of the significance of flares clustering in orbital phase to the expected amplitude of magnetic SPI, and is shown in Figure~\ref{fig:adtest_bp}. We used the custom Anderson-Darling test introduced in Section~\ref{sec:methods:adtest} to look for flaring SPI in all systems with three or more flares with equivalent duration $ED>1\,$s, detected in their Kepler and TESS observations. 

\begin{figure}[ht!]
    \script{paper_adtest_vs_value_scatterplots.py}
    \begin{centering}
        \includegraphics[width=\linewidth]{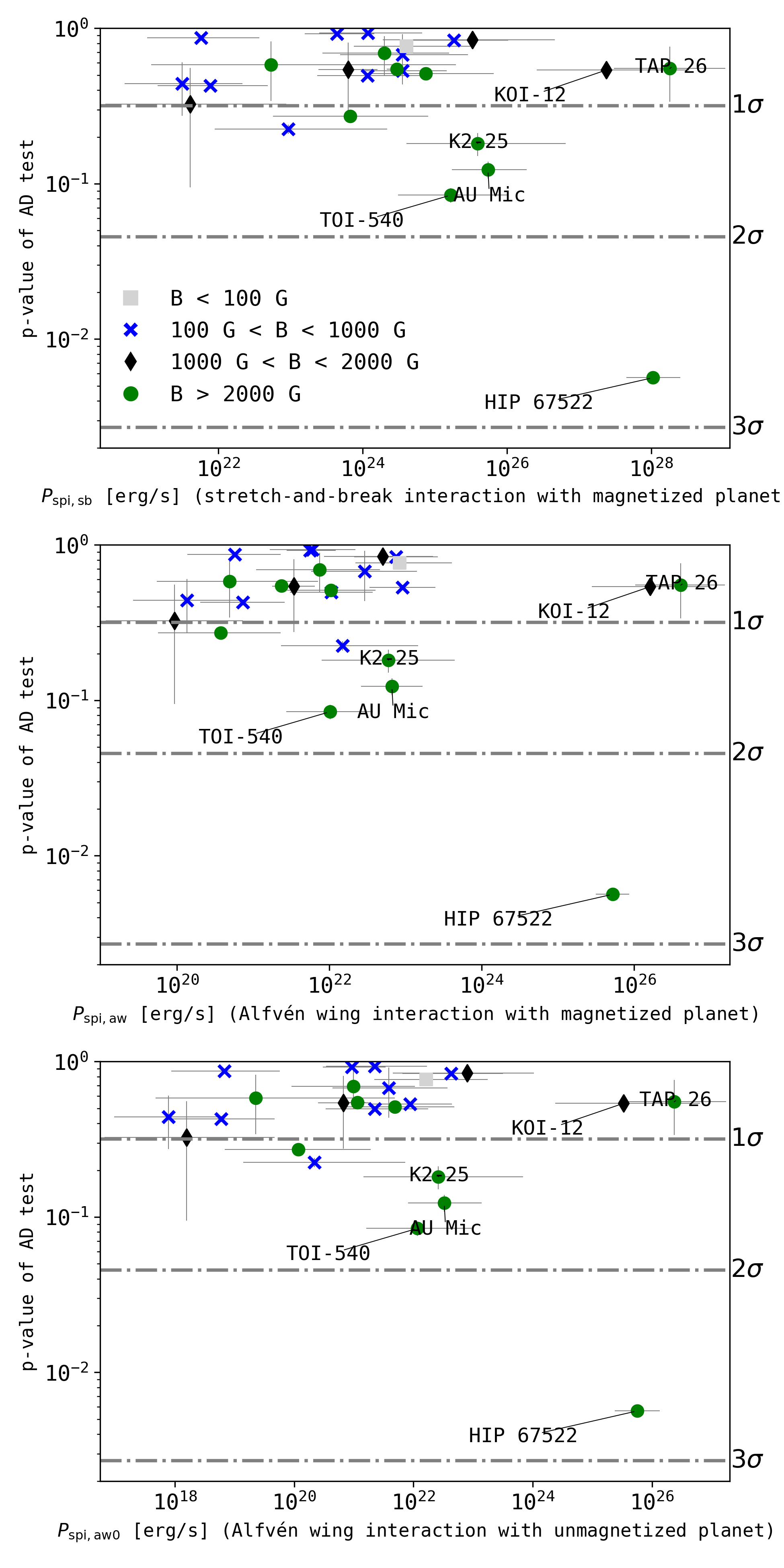}
        \caption{
            Same as Fig.~\ref{fig:adtest_bp}, but for three other scenarios for magnetic SPI. While the distribution of $p$-values is consistent with no interaction, all scenarios indicate lower $p$-values only for high expected powers of interaction. \textbf{Top panel}: stretch-and-break scenario with a magnetized planet. \textbf{Middle panel}: Alfv\'en wing scenario with a magnetized planet. \textbf{Bottom panel}: Alfv\'en wing scenario with an unmagnetized planet. See Fig.~\ref{fig:adtest_ro} in the Appendix for the same figure, color-coded by Rossby number.
        }
        \label{fig:adtest_bp_rest}
    \end{centering}
\end{figure}

Figures~\ref{fig:adtest_bp} and \ref{fig:adtest_bp_rest} show that in every scenario we considered (Alfv\'en wing or stretch-and-break mechanism, with magnetized or unmagnetized planets, see Section~\ref{sec:methods:pspi}), there are two branches on the high end of expected powers. On one branch, the significance of the measured interaction increases with the expected power $P_{\rm spi}$. On the other branch, no interaction is measured regardless of $P_{\rm spi}$. 

Table~\ref{tab:maintable_der} lists the $p$-values for each star-planet system. There is no star-planet system for which we find a $>3\sigma$ ($p<0.0027$) signal of flaring SPI in our sample. However, overall, the significance of the deviation increases with higher expected power $P_{\rm spi}$ of interaction, introduced in Section~\ref{sec:methods:pspi}. Table~\ref{tab:maintable_der} also lists the derived parameters required in these scaling laws, i.e., Rossby number $R$o, surface-average magnetic field strength $B$, and relative velocity $v_{\rm rel}$ between the planet and a co-rotating stellar magnetic field . 

We choose the $ED>1\,$s cutoff to make sure that we are comparing similar flares regardless of spectral type, that is, flares above the same energy relative to stellar luminosity, without losing too many flares. In~\citet{ilin2022searching}, AU Mic appeared to be more modulated with orbital period in the high energy flares above $1\,$s than below. It is also the only star in our sample, for which this threshold makes a difference in the significance of the Anderson-Darling test. However, we note that we lose five systems by applying this threshold -- GJ 393, Kepler-411, Kepler-138, GJ 1061, and Kepler-1084. We note that the flares (with $ED<1\,$s) in all of these systems are consistent with random flare phases within $1\sigma$.

\begin{table*}
\footnotesize
\movetableright=-20mm
\script{paper_main_table.py}
\caption{Flaring star-planet interaction. $R$o, $B$, and $v_{\rm rel}$, are derived from literature values (Table~\ref{tab:maintable_lit}). $P_{\rm spi,xx}$ stands for the power of stretch-and-break ($sb$) and Alfv\'en wing ($aw$) interaction mechanisms, assuming the planet has a magnetic field strength of $1\,$G. $P_{\rm spi,xx0}$ is the same, but assuming an unmagnetized planet. The $p$-value of the Anderson-Darling test is lower when the system shows a stronger clustering of flares in orbital phase. The digits in square brackets refer to the uncertainties in the corresponding last digits shown, e.g., $0.051[1]$ is the same as $0.051\pm0.001$, and $98[12]$ is the same as $98\pm12$.}
\input{output/table_der_vals.tex}
    \label{tab:maintable_der}

\end{table*}

\subsection{Individual systems}
\label{sec:results:individualstars}
The properties of star-planet systems in this work are diverse, including both very fast and very slowly rotating host stars; systems with super-Earths, Neptunes, and Hot Jupiters; and spectral types covering the lower main sequence from mid-F to late M. 

For our analysis, it is important to know that the observed flares indeed occurred on the planet host star. We therefore excluded some systems due to documented contamination from nearby (bound or co-moving) companions~(Section~\ref{sec:results:individualstars:excluded}). One exception to this rule is Kepler-411, a binary system for which the flare contributions from each component could be separated, but which was excluded due to its flares' low $ED$. We also dropped GJ 1061 from the final sample because its rotation period was unconstrained. Another uncertainty in our analysis is the possibility of additional planets at even shorter orbits than the currently known innermost planet, such as might be the case for Proxima Cen~(see Section~\ref{sec:prominensystems}). 

The bulk of the systems in our sample are neither expected to show high power of magnetic SPI, nor do they show any. In Section~\ref{sec:prominensystems}, in the subsample of stars with high expected power, of the order of AU Mic's and above, we first take a closer look at those that seem to follow an increase in measured flaring SPI with increasing expected power $P_{\rm spi}$, that is AU Mic itself, K2-25, TOI-540, and HIP 67522. Then we consider those systems where high power was expected in all scenarios, but not measured: TAP~26 and KOI-12.
\subsubsection{Excluded targets}
\label{sec:results:individualstars:excluded}

While they appear in the flare catalog~(Table~\ref{tab:flares}), we excluded a number of systems from further analysis. We drop all multiple stars except for Proxima Cen (whose components are well separated both physically, and on the sky), and all stars with nearby objects that contaminate our analysis, e.g., from Gaia DR2 and ground-based adaptive optics~\citep{ziegler2018measuring}: \input{output/multiples_string.tex}

For instance, Kepler-808 (KOI-1300) has a $\sim0.4M_\odot$ \citep{kraus2016impact} companion that is about 1.8 mag fainter than the primary at a separation of $0.78\arcsec$~\citep{baranec2016roboao}. TOI-837 has co-moving M dwarf a few arcsec away in the same young cluster, 5 mag fainter, but the primary is an F9-G0 star, so flares could still originate from both stars~\citep{bouma2020cluster}. HD~41004~B, DS~Tuc~A, and LTT~1445 are known multiple systems where all components contribute significantly to the total TESS flux.


\paragraph{Kepler-411}
\label{sec:results:individualstars:kep411}
Kepler-411 has a 3 mag fainter companion~\citep{wang2014influence,ziegler2018measuring}, but we do not exclude it from the analysis initially. The system was observed in Kepler short cadence by \citet{jackman2021stellara}, who found that both the planet host, and the companion star flare. The authors were able to disentangle the contributions from each component on a pixel level. According to \citet{morton2016false, sun2019kepler411}, Kepler-411's planets orbit the primary companion. The fainter companion of Kepler-411 appears to cause the majority of flares (41), whereas the planet host causes only 7. Adopting the 7 flares from \cite{jackman2021stellara} in our analysis, we find that they all have equivalent duration $ED<1\,$s, which excludes Kepler-411 from further analysis. Note, however, that we did not find the low energy flares to cluster in orbital phase, either, after applying our AD test procedure to the 7 correctly attributed flares.

\subsubsection{Prominent systems}
\label{sec:prominensystems}
\paragraph{Proxima Cen}
\label{sec:results:individualstars:proxima}
Although Proxima Cen is part of a triple stellar system, we treat it a single star in this work because of the large angular separation of over 2 deg to its companions, $\alpha$ Cen AB. We do not detect a deviation from intrinsic flaring on Proxima Cen, consistent with numerical models that place Proxima Cen b well outside the sub-Alfv\'enic zone~\citep{alvarado-gomez2020earthlike, kavanagh2021planetinduced, garraffo2022revisiting}. However, this might not entirely rule out flaring SPI in Proxima Cen. The tentatively detected Proxima Cen d, a planet further in, at 0.029 au, or 5 day orbital period~\citep{faria2022candidate, artigau2022linebyline} could still cause flaring SPI. Depending on the phase of the activity cycle, Proxima Cen d may be in the sub-Alfv\'enic regime during Proxima Cen's activity maximum~\citep{alvarado-gomez2020earthlike}. However, the orbital period of the tentative planet is so uncertain at this point that its coherence time~(Section~\ref{sec:results:coherence}) is shorter than the roughly $2\,$yr observing baseline of Proxima Cen, preventing us from accurately measuring the flare phases.

\paragraph{AU Mic, K2-25, and TOI-540}
\label{sec:results:individualstars:aumic}

AU Mic is a 16-29 Myr old pre-main sequence M0-M1 dwarf with a strong magnetic field of about $3010\pm220\,$G obtained from Zeeman broadening measurements~\citep{reiners2022magnetism}. The innermost Neptune-sized planet, AU Mic b, could be inside the sub-Alfv\'enic zone if the star's mass loss rate is relatively low at about 30 times the solar value or lower~\citep{alvarado-gomez2022simulating}. If the mass loss rate is high, exceeding the solar value by a factor of several hundreds, AU Mic b could orbit in the super-Alfv\'enic zone, and become unable to experience any planet-induced flaring. Recently, \citet{klein2022one} found tentative periodicity with the orbital period of AU Mic b in the chromospheric He I D line, which increases if the contribution from flares is included in the calculation. Overall, our marginal signal of flaring SPI in the AU Mic system supports the idea that this young, magnetically active M dwarf system with a close-in Neptune could exhibit planet-induced flaring.
For a detailed discussion of AU Mic and its flaring SPI, we refer to~\cite{ilin2022searching}, where we also estimate that an additional 50–100 days of TESS-like monitoring of AU Mic would yield a $3\sigma$ detection if the marginal signal in our data is real.

K2-25 is a multiplanet system around a fast rotating ($P_{\rm rot}<2\,$d) mid-M dwarf in the Hyades open cluster~\citep[600-800 Myr, ][]{stefansson2020habitable}. Both the expected power of interaction, and the measured deviation, are remarkably similar to the AU Mic system. Interestingly, K2-25 b is in a presumably similar environment to GJ 436 b, but does not show the atmospheric escape in Ly$\alpha$ that the latter famously exhibits, which might be due to high speed stellar winds that suppress atmospheric escape~\citep{rockcliffe2021lya, carolan2020dichotomy, vidotto2020stellar}. As strong winds push the Alfv\'en surface to smaller radii, further investigation of magnetic SPI, flaring or otherwise, in this system could add an important constraint on K2-25's wind properties by clarifying whether the planet orbits in- or outside the Alfv\'en radius, especially accounting for its moderately high eccentricity~(see also Section~\ref{sec:discussion:eccentricity}).

In Figure~\ref{fig:adtest_bp}, K2-25 and AU Mic form a cluster with TOI-540 -- a fast rotating M dwarf with its innermost planet in a $1.24\,$d orbit~\citep{ment2021toi}. Compared to the other two systems, its innermost planet is not a Neptune, but a rocky planet slightly smaller than Earth~\citep{ment2021toi}. It is a relatively smaller obstacle in the host's magnetic field, yet the interaction is expected to be of a similar magnitude due to the strongest inferred stellar magnetic field, and closest orbit among the three systems. 

In the stretch-and-break scenario, and the Alfv\'en wing scenario with a magnetized planet, the three systems cluster together distinctly~(Figure~\ref{fig:adtest_bp}, and top and middle panel in Figure~\ref{fig:adtest_bp_rest}). In the Alfv\'en wing scenario with an unmagnetized planet~(bottom panel in Figure~\ref{fig:adtest_bp_rest}), the distinction is less clear. In both cases, the low level (or possible absence) of interaction could be due to the Alfv\'en radius being within the planetary orbit. It is important to note that the scaling laws introduced in Section~\ref{sec:methods:pspi} assume that the planet is sub-Alfv\'enic, but make no statement about it, because the mass loss of these stars is largely unconstrained. All else equal, a planet is more likely sub-Alfv\'enic if the stellar magnetic field is strong~(see Section~\ref{sec:discussion:as} to see that the three planets are likely sub-Alfv\'enic). Another explanation is the possible intermittency, or 'on-off nature' of the interaction~\citep{shkolnik2008nature}, which may be at play, particularly in cases where the expected power is very strong, such as in the comparison between HIP 67522 and TAP 26 that we address in the following.

\paragraph{HIP 67522}
\label{sec:results:individualstars:hip67522}
has one of the strongest expected SPI signals in our sample, and shows the clearest sign of flaring SPI at $>2.5\sigma$ level ($p<0.006$), with 12 flares in the sample, distributed across three Sectors in TESS. In our original sample from July 2022, HIP 67522 was observed for two Sectors with a total of 6 flares, which yielded a $p$-value of $0.015$. In June 2023, observations from Sector 64 were released, which we used to follow up on this interesting target, increasing the number of flares by another 6, reducing the $p$-value further to $0.0057$. 

HIP 67522 is a young Sun, currently contracting onto the main sequence. It is a Sco-Cen member (10-20 Myr old), which was discovered to host a close-in Jupiter, HIP 67522 b, in 2020~\citep{rizzuto2020tess}. Curiously, the planet is in close spin-orbit commensurability -- $P_{\rm rot}/P_{\rm orb}\approx 1/5$. Therefore, we cannot completely rule out that the observed periodicity is in fact a rotational periodicity, and not related to the planet. The $p$-value of the AD test using the rotational period instead of the orbital period is $p=0.18$. However, the rotational period is very uncertain relative to the total observing baseline (coherence time of about 100 days, and observing baseline of several years), so this number is unreliable.


\paragraph{TAP 26}
\label{sec:results:individualstars:tap26}
is a weak-line T-Tauri star with a strong magnetic field~\citep{yu2017hot}. \citet{lanza2018closeby} estimate that the system can release more energy in magnetic SPI than other  systems with Hot Jupiters, such as HD 179949, that has previously been detected with chromospheric variability in phase with the planet's orbit~\citep{shkolnik2008nature}. However, for TAP 26 b, the orbital period might not be well-constrained. \citet{yu2017hot} apply several methods to derive $P_{\rm orb}$ from the radial velocity data: While the $10.8\,$d orbit we adopted is the most likely according to \citet{yu2017hot}, a $9.0\,$d orbit is also likely, and some of their applied methods favored a $13.4\,$d period. The absence of interaction could hence be a consequence of uncertain orbital period. However, intermittent interactions and the viewing geometry of TAP 26 could also explain the absence of phase-correlated flaring~(see Sections~\ref{sec:discussion:intermittency} and \ref{sec:discussion:viewing}).

\paragraph{KOI-12}
\label{sec:results:individualstars:koi12}
, also known as Kepler-448, is a two-planet system with an eccentric outer planet KOI-12 c, and an inner Warm Jupiter KOI-12 b in a $17.8\,$d orbit~\citep{masuda2017eccentric}. With respect to rotation rate, stellar and planetary radius, the KOI-12 system is similar to HIP~67522 and TAP~26 (although for the those young stars the radii are large because the stars are still contracting onto the main sequence). In contrast, KOI-12 is $1.4\pm0.3\,$Gyr old~\citep{bourrier2015sophie}, an F5 sub-giant~\citep{frasca2016activity}, which is already evolving off of the main sequence. Yet, despite its high expected power of interaction, we detect no excess flaring in phase with KOI-12 b. The reason could be that KOI-12 b is in fact super-Alfv\'enic due to the relatively wide orbit compared to TAP 26 and HIP 67522. Our estimate of the Alfv\'en radius place KOI-12 b in the sub-Alfv\'enic regime~(see Fig.~\ref{fig:as}), but the scaling law applied there may be inaccurate for post-main sequence stars. Intermittent interaction might play a role here as well, and could be favored if future observations capture episodes during which the interaction is 'on'.



\section{Discussion}
\label{sec:discussion}

Our results in Fig.~\ref{fig:adtest_bp} are tentative, yet the trend between expected power of interaction and measured clustering of flares in orbital phase is suggestive, as is the appearance of an active and an inactive branch at high expected powers of magnetic SPI. We can now consider additional facts about the systems, as well as processes, that may explain the observed trends. We consider the extent of the Alfv\'en radius that decides over the possibility of magnetic interaction in the first place~(Section~\ref{sec:discussion:as}), and how orbital eccentricity can further enhance magnetic SPI signal~(Section~\ref{sec:discussion:eccentricity}). The observed branching might also be explained by intermittent SPI~(Section~\ref{sec:discussion:intermittency}), but its possible causes are not well known. As an alternative explanation, we highlight the important influence of viewing geometry on the observability of flaring SPI, which might explain the absence of magnetic SPI in TAP~26~(Section~\ref{sec:discussion:viewing}). 

Magnetic and tidal SPI are difficult to disentangle by their global effects on stellar activity indicators~(Section~\ref{sec:intro:global}). Locally, however, we can discriminate between the two by their relevant periods, that is, $P_{\rm orb}$ and $P_{\rm orb} /2$. In Section~\ref{sec:discussion:tidal}, we repeat the analysis in this paper with $P_{\rm orb}/2$, and find yet more tentative trends in line with scaling laws for tidal interaction. In a system-by-system comparison, we find consistency with the expectation that slowly rotating, likely old, systems can only show local tidal interaction, while rapidly rotating systems with strong magnetic fields are dominated by magnetic forces.
\subsection{Alfv\'en surface}
\label{sec:discussion:as}

\begin{figure}[t]
    \script{paper_as_vs_ad.py}
    \begin{centering}
        \includegraphics[width=\hsize]{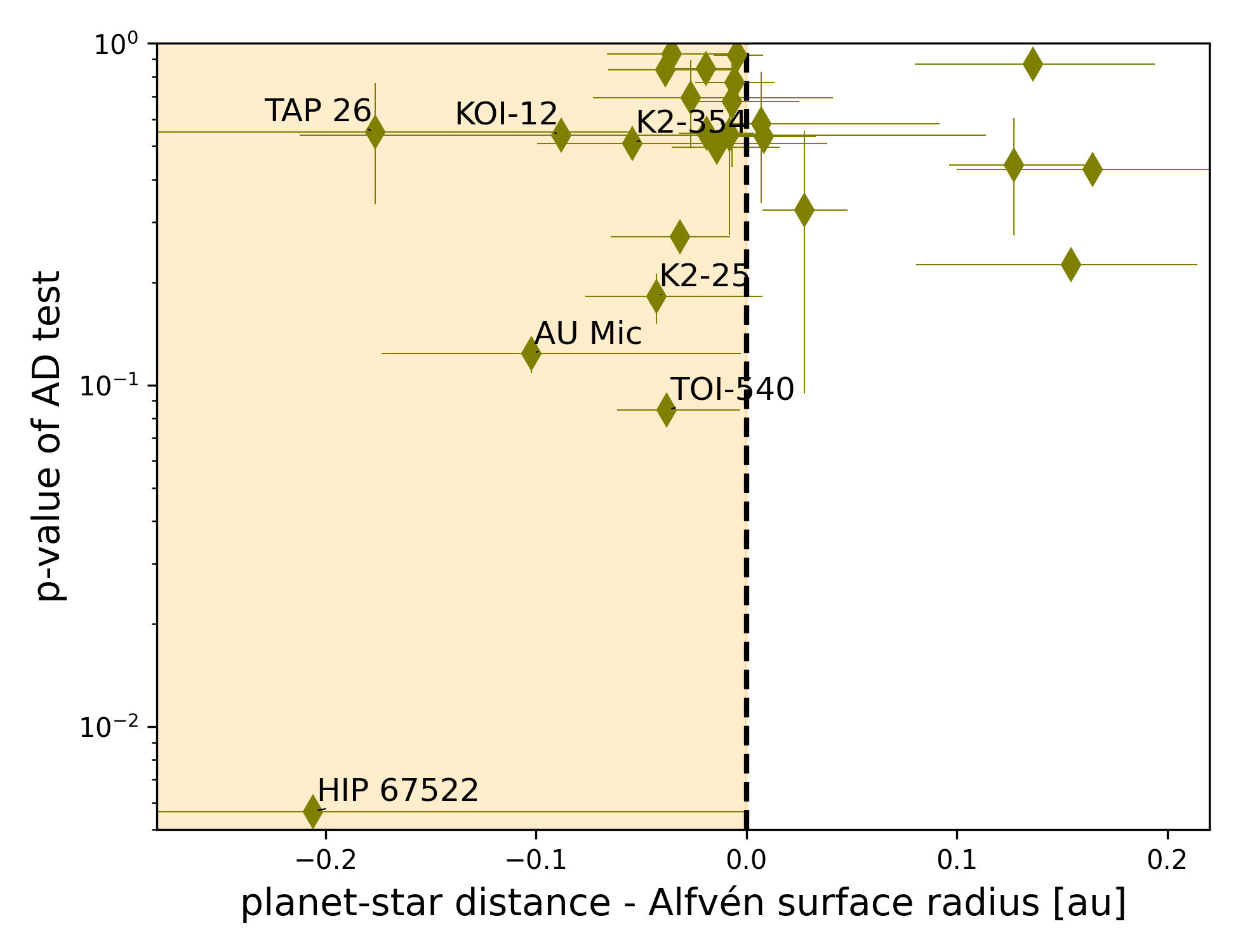}
    \caption{Magnetic flaring star-planet interaction strength ($p$-value of AD test) vs. the innermost planet's location relative to the Alfv\'en surface radius. The innermost planets in the systems left of the dashed line (\textbf{shaded region}) are likely in the sub-Alfv\'enic regime. Only these planets can induce flares in the stellar atmosphere. See Section~\ref{sec:discussion:as} for the procedure we used to estimate the Alfv\'en surface radius. }
        \label{fig:as}
    \end{centering}
\end{figure}

Magnetic SPI can only take place if the planet spends at least a part of its orbit in the sub-Alfv\'enic regime~(see Section~\ref{sec:intro:mspi}). We estimated the average Alfv\'en surface size ($\overline{AS}_{\rm R}$) using the power law established by \citet{chebly2023numerical}:

\begin{equation}
    \log \overline{AS_{\rm R}} = (0.44 \pm 0.05) \log B_{\rm r}^{\rm max} + (0.54 \pm 0.08)\label{AS_estimation}
\end{equation}

This relation links the mean AS size with the maximum large-scale radial magnetic field ($B_{\rm r}^{\rm max}$) obtained in Zeeman-Doppler Imaging reconstructions (\citealt{donati2006surprising, morin2008largescale, fares2009magnetic, alvarado-gomez2015activity, hussain2016spectropolarimetric, kochukhov2020mapping}). The relation was derived from 3D~MHD stellar wind simulations of cool main sequence stars (F, G, K, and M), which used observed ZDI maps as boundary condition for the stellar radial magnetic field (for more details on the models see \citealt{chebly2023numerical}). However, the stellar magnetic field~($B$, Table~\ref{tab:maintable_der}), derived from the Rossby number (Eq.~\ref{Eq3}, \ref{Eq4}), corresponds to the total, unsigned magnetic field strength on the stellar surface so that it is necessary to estimate how $B_{\rm r}^{\rm max}$ is related to $B$. 

It is important to note that \citet{chebly2023numerical} assume a consistent Poynting flux per unit magnetic field strength for all stars, using the value of the Sun as a reference. The size of the Alfv\'en surface depends strongly on this assumption. As such, any variation in the value between different stars would lead to deviations from the law proposed in \citet{chebly2023numerical}, and change the dimensions of the Alfv\'en surface. Nevertheless, while the consistency in the Poynting flux appears as a very strong assumption, the numerical results of \citet{chebly2023numerical} fare relatively well when compared with our current observational knowledge of stellar winds in cool main-sequence stars~\citep{wood2021new}. For this reason, we consider that Eq.~\ref{AS_estimation} should be robust enough for our purposes.

\citet{reiners2022magnetism} report that for Sun-like stars with a total magnetic field strength of up to 2000~G, the longitudinal magnetic field derived from Stokes V spectropolarimetric observations ($B_{\rm V}$) accounts for $\sim$\,$10$\% of the total magnetic field. \cite{kochukhov2020hidden} showed that this trend also holds for M-dwarfs with magnetic fields up to 4000~G. In this case, however, the recovered longitudinal magnetic field accounts for about $20\%$ of the total magnetic field. Moreover, our star sample covers a wide range of magnetic fields, the weakest being 100~G and the strongest $3800\,$G. Therefore, as a first assumption, we consider that the longitudinal magnetic field of our stars would reach 15\% of the total magnetic field (i.e. $B_{\rm V} = 0.15B$), given that most systems in our sample have $B>2000\,$G. 

The second assumption is that $B_{\rm V}$ will match the maximum radial magnetic field $B_{\rm r}^{\rm max}$ in a ZDI reconstruction. $B_{\rm V}$ is integrated over the visible surface of the star. Only if the effect from cancellation of opposite polarities is small for an observing phase that contains the strongest large-scale radial magnetic region, will this assumption hold. At all other rotational phases, we expect $B_{\rm V} < B_{\rm r}^{\rm max}$. Therefore, by assuming $B_{\rm V} = B_{\rm r}^{\rm max}$ we are effectively taking an upper limit on the longitudinal magnetic field strength, which will be translated\footnote[3]{Naturally this will also depend on how good our $15$\% match between $B$ and $B_{\rm V}$ is to reality.} in our estimated values of $\overline{AS}_{\rm R}$.

We show the difference between star-planet distance and $\overline{AS}_{\rm R}$ in Fig.~\ref{fig:as}. HIP 67522 b falls most deeply into the sub-Alfv\'enic regime, consistent with the strongest signal of magnetic SPI in our sample. K2-25, TOI-540 and AU Mic also fulfil the condition of sub-Alfv\'enic orbit of the innermost planet, consistent with their observed marginal deviation from uniform flaring with orbital phase. 

TAP 26 and KOI-12 also appear sub-Alfv\'enic, despite absent signal of magnetic SPI. From this follows that the absence of interaction is not caused by the planets orbiting outside the Alfv\'en radius. Location inside $\overline{AS}_{\rm R}$ is a required but not sufficient condition for magnetic SPI, as we will see in the following subsections. 

\subsection{Eccentricity}
\label{sec:discussion:eccentricity}
Orbital eccentricity can affect magnetic SPI by modulating the planet's distance to the star. If the planetary orbit is highly eccentric, the difference between periastron and apoastron can be significantly larger than the variability of the Alfv\'en radius. If then apoastron is outside this radius, and periastron within, the planet may find itself in the sub-Alfv\'enic zone only briefly, during the planet's rapid periastron passage. This was previously exploited in the case of HD~17156~\citep{maggio2015coordinated}, and analogously for the colliding magnetospheres in binary systems~\citep{massi2002periodic,getman2016search,das2023discovery}. In our sample, the estimates for eccentricities of the innermost planets are all moderate to low, with $e\leq 0.35$. The exception is K2-25, with a relatively high $e\approx0.43$. If periastron passage in K2-25 occurs when the planet is in front of the star relative to the observer, the visibility effect of the interaction footpoint is further increased by the narrow phase range of the periastron passage. \citet{stefansson2020habitable} estimated the argument of periastron $\omega=120^{+12}_{-14}\,$deg for K2-25 (with longitude of the ascending node chosen such that $\omega$ is the same as the longitude of periastron, with $\omega=90\,$deg being periastron passage during transit, see~\citealt{kipping2010investigations,dawson2012photoeccentric}). We can therefore conjecture that the tentative signal of flaring SPI in K2-25 is enhanced by eccentricity. Statistically, we are more likely to observe a planetary transit close to periastron, since periastron permits more orbital inclinations to transit than apoastron. Overall, we expect eccentric transiting systems to experience elevated SPI signal, which could be tested in the future when better constraints on eccentricity become available~(see Table~\ref{tab:maintable_lit}).

\subsection{Intermittency}
\label{sec:discussion:intermittency}
Magnetic SPI power in a star-planet system may sometimes be low for episodes of time, and high for others. HD~179949 is the prototype example for this phenomenon. The orbital modulation of its chromospheric indicators was observable for four out of six epochs on a five-year baseline~\citep{shkolnik2003evidence,shkolnik2008nature}. It is not clear how long a continuous epoch of interaction or non-interaction might be. This leaves room for many different explanations. Candidate mechanisms for the star include: 
\begin{itemize}
    \item the Alfv\'en radius moving in- and out the planet's orbit, either over the course of an activity cycle, or through short-term variations in the magnetic field and wind properties caused by, e.g., coronal mass ejections;
    \item the interacting footpoint moving to higher latitudes due to changes in the large scale field, which reduces the modulation of its visibility; 
    \item an active latitude (i.e., small scale field) moving away from the magnetic footpoints passageway;
    \item the available magnetic energy decreasing such that the flares or other magnetic interaction indicators still occur but fall below the detection threshold for a period of time; or
    \item the availability of a sufficient amount of plasma in the vicinity of the magnetic interaction region to yield significant observable flare emission in the stellar atmosphere (in analogy to the Jupiter-Io system, wherein Io's volcanism provides the plasma that eventually produces an interaction hotspot on Jupiter,~\citealt{clarke1996farultraviolet, prange1996rapid})
\end{itemize}

If the magnetic field of the planet also changes over time, this might modulate the intensity as well~\citep[e.g.][]{turnpenney2018exoplanetinduced}.

The location of active regions within the passageway of the magnetic footpoints may be such that the interaction occurs only when the region faces away from the observer. This could span multiple orbits of the planet, particularly when the surface is populated by few stable active regions, and the orbital and rotational periods are similar. The flip-flop behavior of active longitudes seen in young solar-like stars~\citep{berdyugina2005starspots} could then cause a relatively sudden quenching of interaction by moving the interacting region out of view for a period of time. 

In our results in Fig.~\ref{fig:adtest_bp}, we can interpret the two branches emerging at high expected power of interaction as due to intermittency. TAP 26 and KOI-12 do not show any sign of excess flaring in our data, despite similar expected powers as in HIP 67522 with its $>2.5\sigma$ signal. However, other explanations cannot be ruled out yet (see Section~\ref{sec:results:individualstars}, and the following).


\subsection{Viewing geometry}
\label{sec:discussion:viewing}
Modulated visibility of the magnetic interaction footpoint is crucial for the detection of flaring SPI in our analysis~(Fig.~\ref{fig:sketch}). The viewing geometry of the star-planet system is therefore important to consider as an alternative to, or explanation for, the intermittency discussed above. Assuming for the sake of the argument that the sub-planetary point is close to the magnetic interaction footpoint in longitude, two factors will determine whether the footpoint's visibility will be modulated with the planetary orbit or not: the orbital inclination and the footpoint's latitude relative to it. 

Consider a system, where orbit, and stellar spin and stellar magnetic dipole are aligned~(Fig.~\ref{fig:sketch}). If the magnetic footpoint's latitude is low, i.e., close to the orbital plane, the orbital plane can be inclined to a high degree, and the footpoint will still periodically move in and out of view. However, if the footpoint's latitude is high, we need to observe the star-planet system nearly edge-on to be able to measure a modulation of activity with the planet's orbital phase. Assuming the dipole field dominates at the distance of the planet, the footpoint will be close to the pole, similar to the UV spot created by Io and the other moons in the polar cap of Jupiter~\citep{clarke1996farultraviolet, prange1996rapid}. In this scenario, many, especially non-transiting, systems may experience flaring SPI, but those would not show up as correlation of flare timing with the orbital phase in our data.

Most of our systems are transiting. For those we can expect this scenario to be a minor problem unless the latitude of the footpoint is very high. Of the systems that have high expected power but do not show any excess flaring, only TAP 26 has a planet detected in radial velocity with a relatively high orbital inclination of about $55\pm10$ deg inferred from photometry, spectroscopy, and Zeeman Doppler measurements in~\citet{yu2017hot}. Their ZD map reveals magnetic field concentrations close to the rotational pole, and the brightness maps also show a dark spot in the same location. If the orbital plane axis is aligned with the stellar rotation axis, the absence of phase-correlated flaring in TAP 26 could be explained by the footpoint of interaction being always in view, and therefore never modulated. 

However, alignment between orbital, rotational and magnetic dipole axis is not generally given. Many planets orbit their hosts nearly pole-on~\citep{albrecht2012obliquities, albrecht2022stellar, bourrier2023dream}. T Tauri stars show somewhat misaligned magnetic axes relative to their spin~\citep{mcginnis2020magnetic}. In Sun-like stars, the magnetic dipole axis location changes throughout the activity cycle~\citep{petit2009polarity,borosaikia2018direct}, which may also cause temporary cessation of visibility. Orbital precession of misaligned planets can also cause such intermittency, if it periodically brings the footpoints trajectory on the stellar surface fully into view so that it is no longer modulated.

\subsection{Tidal interaction}
\label{sec:discussion:tidal}
\begin{figure*}[t]
    \script{paper_tidal_interaction.py}
    \begin{centering}
        \includegraphics[width=\hsize]{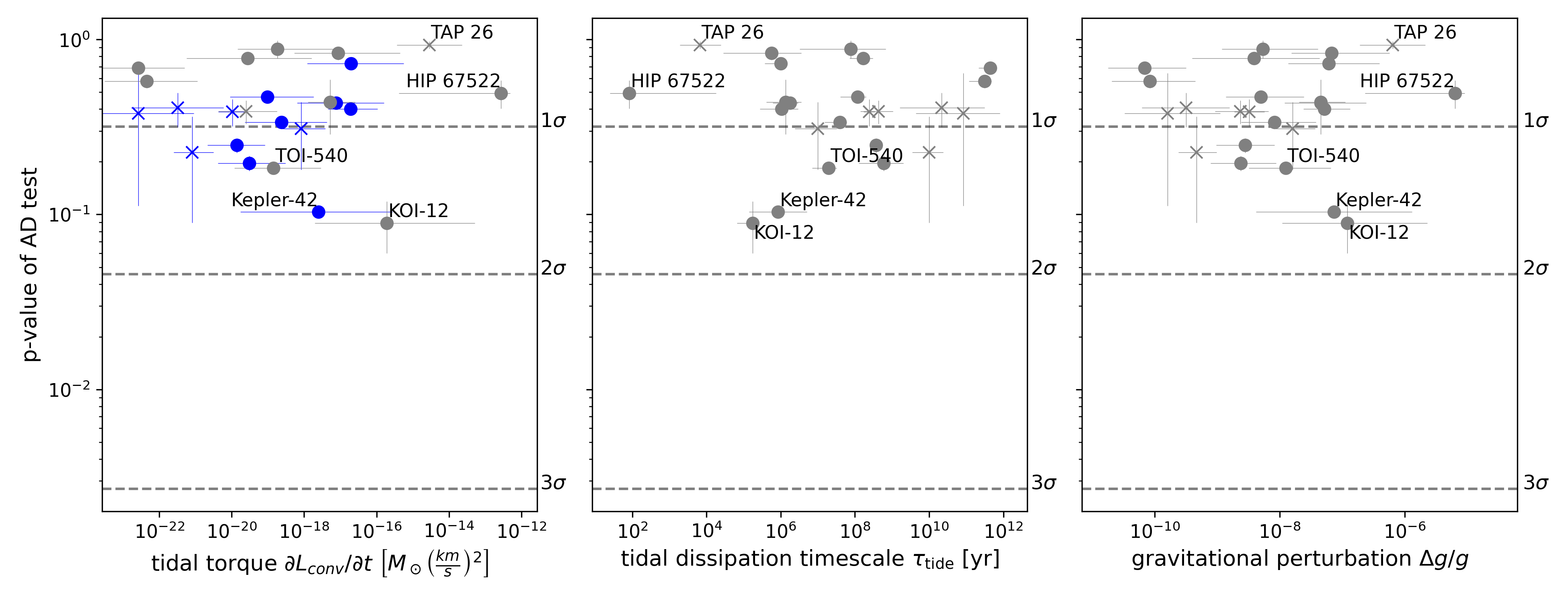}
    \caption{Deviation from random flare times in phase with $P_{\rm orb}/2$, i.e., in phase with the two tidal bulges raised on the star, compared to different models of tidal interaction~(Section~\ref{sec:discussion:tidal}). \textbf{Cross symbols} mark planets with $M_{\rm p} \sin i$ measurements. \textbf{Round symbols} mark planets with measured  or estimated $M_{\rm p}$ (see Section~\ref{sec:discussion:mpms}). 
           \textbf{Left panel}: Tidal torque in the system for stars with a convective envelope. Blue color indicates that the star rotates slower than the planet orbits, so that angular momentum is transferred from the orbit to the spin of the star. Grey color is the reverse. \textbf{Middle panel}: The tidal dissipation timescale. \textbf{Right panel}:  Relative gravitational perturbation in the star.}
        \label{fig:tidal}
    \end{centering}
\end{figure*}

\begin{table*}
\script{paper_table_tidal.py}
\footnotesize
\movetableright=-20mm
\caption{Flaring tidal star-planet interaction parameters and $p$-value of the custom Anderson-Darling test. }
\input{output/table_tidal.tex}
    \label{tab:tidal}
 \tablerefs{\input{output/table_tidal_bibstring.tex}}
\end{table*}

\subsubsection{Locally expressed tidal interaction}
Tidal interaction can also lead to enhanced stellar activity, globally, and locally, e.g., higher coronal emission~\citep{ilic2022tidal}, or excess flaring. Locally, we expect to see a periodicity of activity with half the orbital period of the planet, corresponding to the two tidal bulges forming in the stellar envelope~\citep{cuntz2000stellar}. \citet{holzwarth2003dynamics} suggest an asymmetry induced in the stellar dynamo by the tidal disturbance, which could lead to a preference for active regions (potentially flaring) at longitudes about $P_{\rm orb}/2$ apart from each other. 

No observations of local tidal SPI exist yet. However, in the extreme case of RS CVn binaries, starspots often appear modulated with half the orbital period of the system~\citep{olah2002time, ozavci2018recurrent, kriskovics2023ei}. A similar effect was measured when the companion was likely a very low mass star or a brown dwarf~\citep{donati1995activity, frasca2008spots, parks2021interferometric}, and in the case of a white dwarf companion~\citep{hussain2006spot, watson2007roche}. In ellipsoidal binaries, a modulation of flares with $P_{\rm orb}/2$ beyond geometrical effects was observed with Kepler~\citep{gao2016whitelight}, suggesting flaring regions at the location of the extreme tidal bulges in these stars. However, all these systems are synchronized, $P_{\rm rot}=P_{\rm orb}$, so that intrinsic active longitudes~\citep{usoskin2007longterm, weber2013theory, jarvinen2005spots, lanza2009corot} cannot be disambiguated from the result of tidal star-star interaction ~\citep{holzwarth2003dynamics}. 

If it is the stellar companion that tidally causes the observed effects, planetary companions should do the same, albeit at a smaller magnitude. Fortunately, spin and orbit are not synchronized in our sample, so confusion with intrinsic effects is less likely. 

\subsubsection{Tidal interaction scaling laws}

We cannot tell whether an additional individual flare is triggered by tidal or magnetic interaction, but we can look for deviations from a random distribution of flare times in phase with the relevant period, i.e., $P_{\rm orb}/2$. 

We applied the same technique as described in Section~\ref{sec:methods:adtest} to determine the significance of a local tidal SPI signal. Since we lack a model to relate tidally induced flaring to the systems' properties, we follow~\citet{ilic2022tidal}, and compare the resulting $p$-values with three different scaling laws for tidal interaction as proxies for expected excess flaring due to tides in Fig.~\ref{fig:tidal}. We list the derived values in Table~\ref{tab:tidal}. 

The first model measures the torque $\partial L_{\rm conv} / \partial t$, exerted by angular momentum transfer between star and planet, assuming the star has a convective envelope~(\citealt{penev2012constraining}, their Eqns.~1~and~2). We choose the tidal quality factor $Q_*=10^7$, i.e., low efficiency of tidal dissipation, as a conservative estimate. Our results in Fig~\ref{fig:tidal} (left panel) do not depend on the choice of $Q_*$, as we only need to know the tidal torque down to a proportionality factor. We also make the simplifying assumption that the rotation period of the convective envelope is the same as the surface rotation period measured as described in Section~\ref{sec:data:rotationperiods}, ignoring differential rotation effects.

The second model considers the tidal dissipation timescale -- the shorter the timescale, the higher the power of the interaction~\citep{zahn1977tidal}. Following~\citet{ilic2022tidal}, we adopt Eq.~2 in~\citet{albrecht2012obliquities} for the tidal dissipation timescale in the convective envelope. We assume that all stars in our sample have a convective envelope because they flare, and flaring requires a dynamo-generated magnetic field, which can only operate in stars with convective envelopes or fully convective stars. Most of our stars are dwarf stars below the Kraft break at spectral type F5~\citep{kraft1967studies}, except for KOI-12, which is an F5 sub-giant~\citep{frasca2016activity}.

The third model uses the gravitational perturbation as a proxy for the interaction~\citep{cuntz2000stellar}, originally derived for SPI. Their Eq.~1 proposes a gravitational perturbation proportional to the mass ratio between planet and star, and $a^{-3}$. 

\subsubsection{Planetary and stellar masses}
\label{sec:discussion:mpms}
We collect the planetary and stellar masses from the literature~(Table~\ref{tab:tidal}). For missing planetary masses, the radii were known, so we used the mass-radius relations implemented in \texttt{astro-forecaster}~\citep[Ben Cassesse's implementation of \texttt{forecaster},][]{chen2017probabilistic}. GJ 3323, GJ 674 and GJ 3082 were also missing stellar mass estimates, which we obtained from the \citep{mann2015how, mann2016erratum} relations between absolute $K_{\rm s}$ magnitude~\citep{skrutskie2006two} and stellar mass for M dwarfs using the distances from \citet{bailer-jones2021estimating}. For Kepler-42 b, we used  $M_{\rm p}=0.1-2.06M_\oplus$ that covers pure rock to pure iron compositions in \citet{muirhead2012characterizing}, and adopted the logarithmic mean for our estimate. Kepler-1558 c has a similar radius to Kepler-42 b, i.e., most likely a bare core planet, so we used density ranging from Earth bulk density to pure rock to estimate its mass. We also adopt the upper limit of $5M_{\rm J}$ for HIP 67522 from~\citet{rizzuto2020tess}, but use the lower error from \citet{chen2017probabilistic}. The assumptions we made here are conservative, and the estimates based on \citet{chen2017probabilistic} use a broad sample of planets, so we consider the planetary mass estimates with their respective uncertainties robust. We note, however, that the mass estimates for RV-detected planets are $M_{\rm p} \sin i$, so that their expected interaction may be stronger than indicated in Fig.~\ref{fig:tidal}. 

\subsubsection{Tidal vs. magnetic flaring star planet interaction}

We find that no single system shows significant signs of tidal SPI. Overall, the signal in phase with $P_{\rm orb}/2$ is smaller than with $P_{\rm orb}$. However, in all three scenarios, the sample's deviation from random flare timing increases with expected tidal interaction strength. We can compare the magnetic interaction measurements to the tidal ones in some of the conspicuous systems:

\paragraph{HIP 67522} does not show signs of tidal interaction, despite a relatively high expected power. Conceivably, both effects play a role at the same time, but the tidal interaction is dominated by magnetic SPI. With more available observations in the future, and better constraints on the expected signal of each type of SPI, it will become possible to model both simultaneously. However, with only a total of 12 flares in this work, we cannot make any strong claims about the contribution of tidal SPI, except for it apparently being weaker than the magnetic aspect.  

\paragraph{KOI-12,} which lacks signal with $P_{\rm orb}$ despite high expected power of magnetic interaction, shows elevated expected and measured tidal interaction. However, we note that KOI-12 may be evolving off the main sequence, so that the tidal interaction scaling laws may not apply for this star. 

\paragraph{TAP 26,} which shows no signs of magnetic SPI, does not show any sign of tidal interaction, either, despite both being high according to the scaling laws. This could again be due to the inclination of TAP 26's orbit, but only if the flares caused by the tidal bulge are not close to the equator. Otherwise, TAP 26's tidally induced flares should be more modulated than the magnetically induced ones. 

\paragraph{Kepler-42 and TOI-540} Kepler-42 is neither expected to interact magnetically, nor did it show any deviation from random flare timing in phase with $P_{\rm orb}$. However, since it is the system with the shortest planetary orbit in our sample, it is both expected to and does show marginal signal of tidal interaction. Kepler-42 is very similar to TOI-540, which also is a small M dwarf with a likely terrestrial planet in a very short orbit. A key difference is that Kepler-42 is rotating at about 70 days, and therefore expected to have a relatively weak magnetic field, whereas TOI-540 is a young star with $P_{\rm rot}<1\,$d. TOI-540 is expected to have lower tidal interaction strength than Kepler-42 due to TOI-540 b's wider orbit, and also shows a weaker tidal flaring SPI signal in all three scenarios. 

\paragraph{K2-25 and AU Mic} cluster with TOI-540 in magnetic SPI -- all three show tentative signs of magnetic interaction and similar expected powers. But in contrast to TOI-540, both K2-25 and AU Mic are consistent within $1\sigma$ with no observable tidal interaction.

Overall, tidal interaction, if at all present, is less pronounced than magnetic SPI in our sample. However, if the observed deviations from intrinsic flaring do in fact represent low level tidal interaction signal, we can conclude that close-in planets can interact both magnetically and tidally. In line with theoretical considerations~\citep{strugarek2017fate}, our results tentatively suggest that magnetic interaction is more pronounced for planets with fast rotating, active hosts, while tidal interactions dominate systems with slowly rotating, inactive stars. Finally, exceptions such as TAP 26 exemplify that the simple scaling laws we applied here may not capture the complexity of the interaction.

\section{Summary and Conclusions}
\label{sec:summary}

We conducted the so far largest search for flaring star-planet interactions (SPI). Using over 7200 Quarters and Sectors of photometric monitoring from Kepler and TESS archives, we searched for flares in over 1800 systems of the about 3000 listed in the NASA Exoplanet Archive. When flares could be unambiguously attributed to the planet host, and showed three or more energetic events in the data, we searched for flares that clustered in orbital phase. Our final sample consisted of 25 systems, among them well-known flaring hosts like Proxima~Cen, AU~Mic, and TRAPPIST-1. 

Applying both the stretch-and-break and Alfv\'en wing mechanisms for magnetic star-planet interaction, we found a tentative trend between the expected power of interaction, and the presence of excess flaring in phase with the orbital period $P_{\rm orb}$. In particular, we found that there may be two branches at high expected power -- one where the measured orbital clustering increases with power, and one without a trend. While it is unlikely that the extent of the Alfv\'en surface is a reason, intermittency in the interaction and the viewing geometry of the systems could explain the branching. Among the investigated systems, the flares in HIP~67522, a young Sun-like Hot Jupiter host, cluster most distinctly in orbital phase, consistent with its deep embedding in the sub-Alfv\'enic zone and its high expected power of interaction.

We also searched the same data for signs of locally expressed tidal interaction, that is, excess flaring in phase with the tidal bulges ($P_{\rm orb}/2$). We found a similar trend with different scaling laws of tidal interaction, albeit at even lower significance. Our system-by-system comparison suggests that young systems, like HIP~67522, may be dominated by magnetic interaction, whereas old, magnetically inactive systems, like Kepler-42, might mostly interact tidally. 

Our study is the first that systematically searched for flaring SPI in a large sample of systems with excellent phase coverage. While our results remain tentative, future studies using our technique will benefit from accumulating data. The TESS mission continues to produce high precision light curves for many nearby active planet hosts, and the PLATO~\citep{rauer2014plato} mission will join in late 2026. For AU Mic, we may be only a few TESS Sectors away from a $3\sigma$ level confirmation~\citep{ilin2022searching} of the tentative interaction signal observed here and in other work~\citep{klein2022one}, and similarly for HIP 67522. Several systems in our sample are already scheduled for further observing with TESS, including Kepler-42, and KOI-12.


\section*{Acknowledgements}
The authors thank the anonymous referee for their thorough criticism of the manuscript, and numerous valuable suggestions. EI would like to thank Antonino Lanza for helpful advice.
This paper includes data collected by the TESS mission, which are publicly available from the Mikulski Archive for Space Telescopes (MAST).
Funding for the TESS mission is provided by NASA’s Science Mission directorate. 
This work has made use of data from the European Space Agency (ESA) mission {\it Gaia} (\url{https://www.cosmos.esa.int/gaia}), processed by the {\it Gaia} Data Processing and Analysis Consortium (DPAC, \url{https://www.cosmos.esa.int/web/gaia/dpac/consortium}). Funding for the DPAC
has been provided by national institutions, in particular the institutions participating in the {\it Gaia} Multilateral Agreement.
KP, JC and NI acknowledge funding from the German \textit{Leibniz Community} under grant P67/2018.
\section*{Data Availability}
All data used in this study are publicly available in their respective archives.
This study used the reproducibility software \href{https://github.com/showyourwork/showyourwork}{showyourwork}
\citep{luger2021mappinga} to create all figures, format all tables, and compile the manuscript. Each figure links to the script that produced the figure in the git repository~\href{https://github.com/ekaterinailin/flaring-spi-paper}{github.com/ekaterinailin/flaring-spi-paper}. The almost 13000 de-trended light curves, and the full versions of all Tables in this work are available via Zenodo \href{https://doi.org/10.5281/zenodo.8355002}{10.5281/zenodo.8355002}. The git repository that produces the data stored on Zenodo is publicly available at \href{https://github.com/ekaterinailin/flaring-spi}{github.com/ekaterinailin/flaring-spi}.
\bibliography{bib}

\appendix
\restartappendixnumbering

\section{Expected power vs. AD test, color-coded by Rossby number}
\begin{figure*}[ht!]
    \script{paper_adtest_vs_value_scatterplots.py}
    \begin{centering}
        \includegraphics[width=\linewidth]{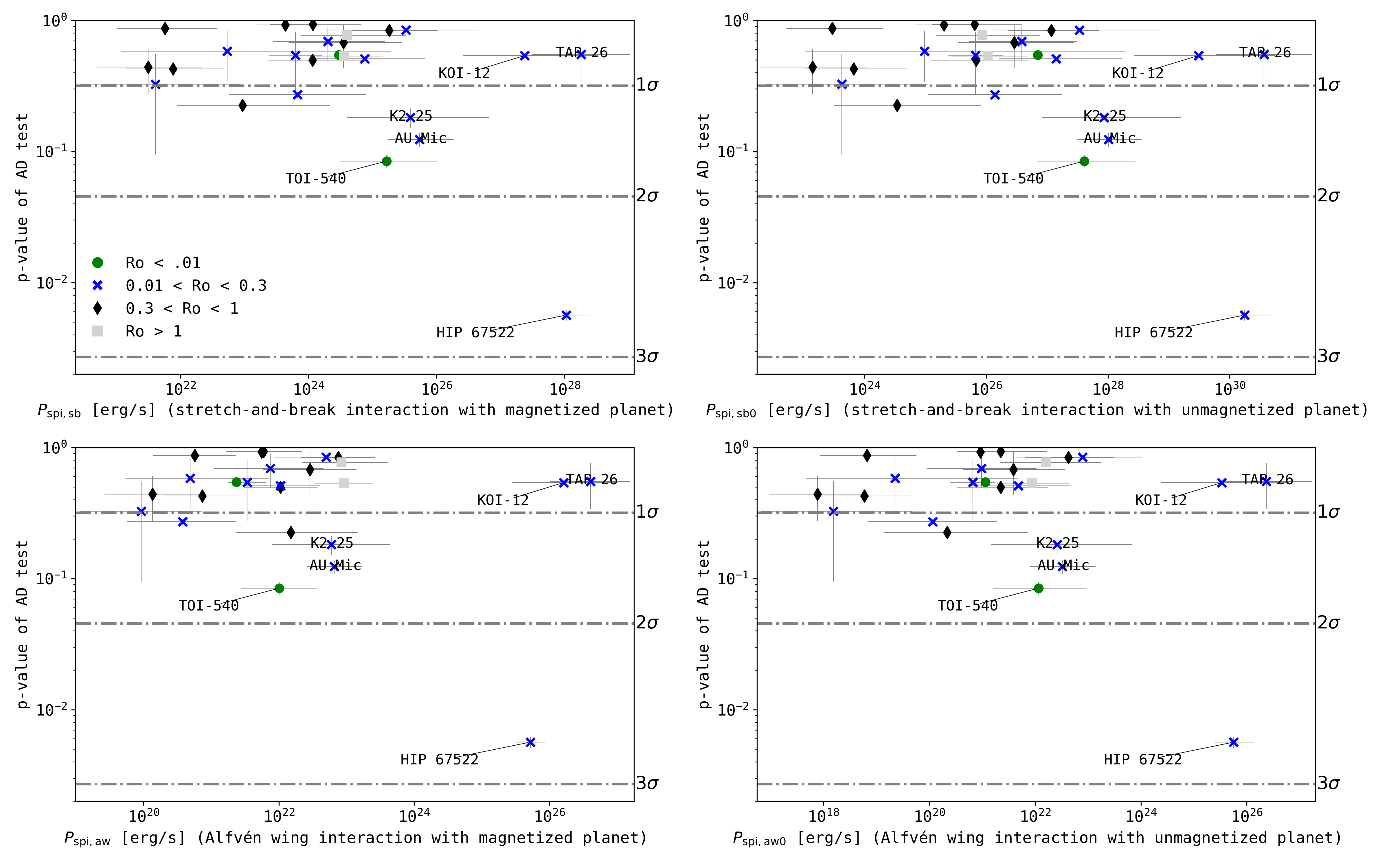}
        \caption{
        Expected power of flaring SPI vs. AD test results, assuming the same four scenarios as in Figures~\ref{fig:adtest_bp} and \ref{fig:adtest_bp_rest}, here color-coded by Rossby number. $R\rm o=0.3$ is chosen to mark the transition from the saturated ($R\rm o < 0.3$) to the unsaturated ($R\rm o > 0.3$) activity regime. 
        }
        \label{fig:adtest_ro}
    \end{centering}
\end{figure*}

\newpage
\section{Finding flares in Kepler and TESS: False positives}
\restartappendixnumbering 

\begin{figure*}[ht!]
    \script{paper_false_positives.py}
    \begin{centering}
    \includegraphics[width=\linewidth]{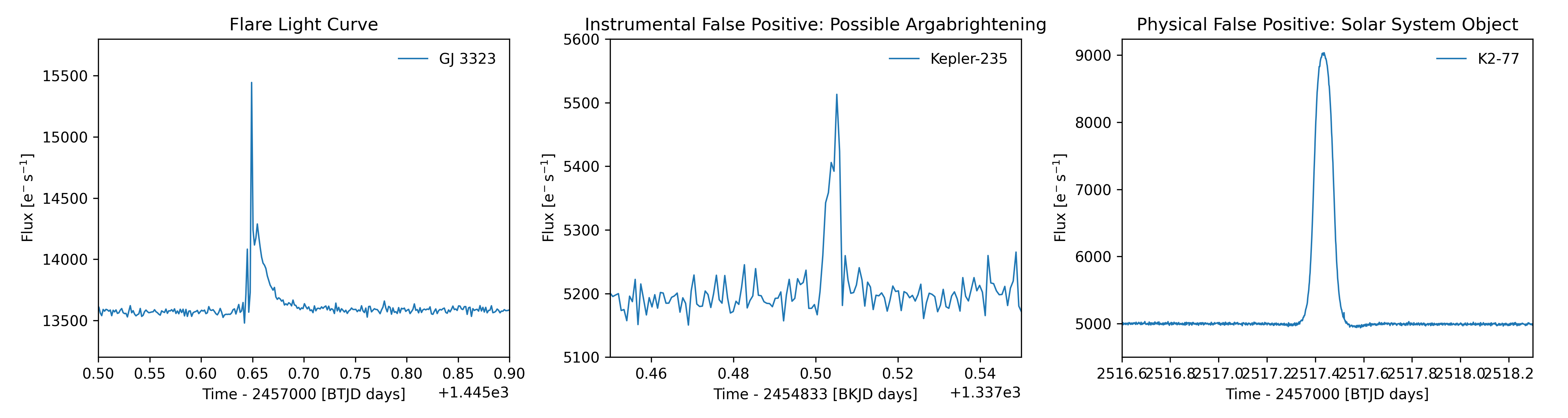}
    \caption{True and false positive flare candidates in Kepler and TESS light curves. \textbf{Left panel}: A flare in the TESS 2-min cadence light curve of GJ 3323 in Sector 5, with its characteristic fast-rise-exponential-decay shape, and additional substructure around the peak. \textbf{Middle panel}: Kepler-235 with a possible argabrightening in the 1-min cadence light curve in  Quarter 14. Argabrightenings are few-minutes long reflections of light off of debris and onto the detector~\citep{vancleve2016kepler}. The reflections affect large parts of the detector, so that this event can be identified as a false positive by its common occurrence across many light curves. \textbf{Right panel:} K2-77 with a Solar System Object (SSO) moving across the TESS detector in the 2-min cadence light curve in Sector 44. SSOs produce smooth and nearly symmetrical brightenings as they move in and out of the aperture of the star. They can also (partially) be identified with known SSOs~\citep{berthier2006skybot, berthier2016prediction}.}
    \label{fig:flares_and_fps}
        \end{centering}
\end{figure*}

\section{Anderson-Darling $A^2$ statistic}
\restartappendixnumbering 

\begin{figure*}[ht!]
    \script{paper_example_ad_dist.py}
    \begin{centering}
    \includegraphics[width=0.5\linewidth]{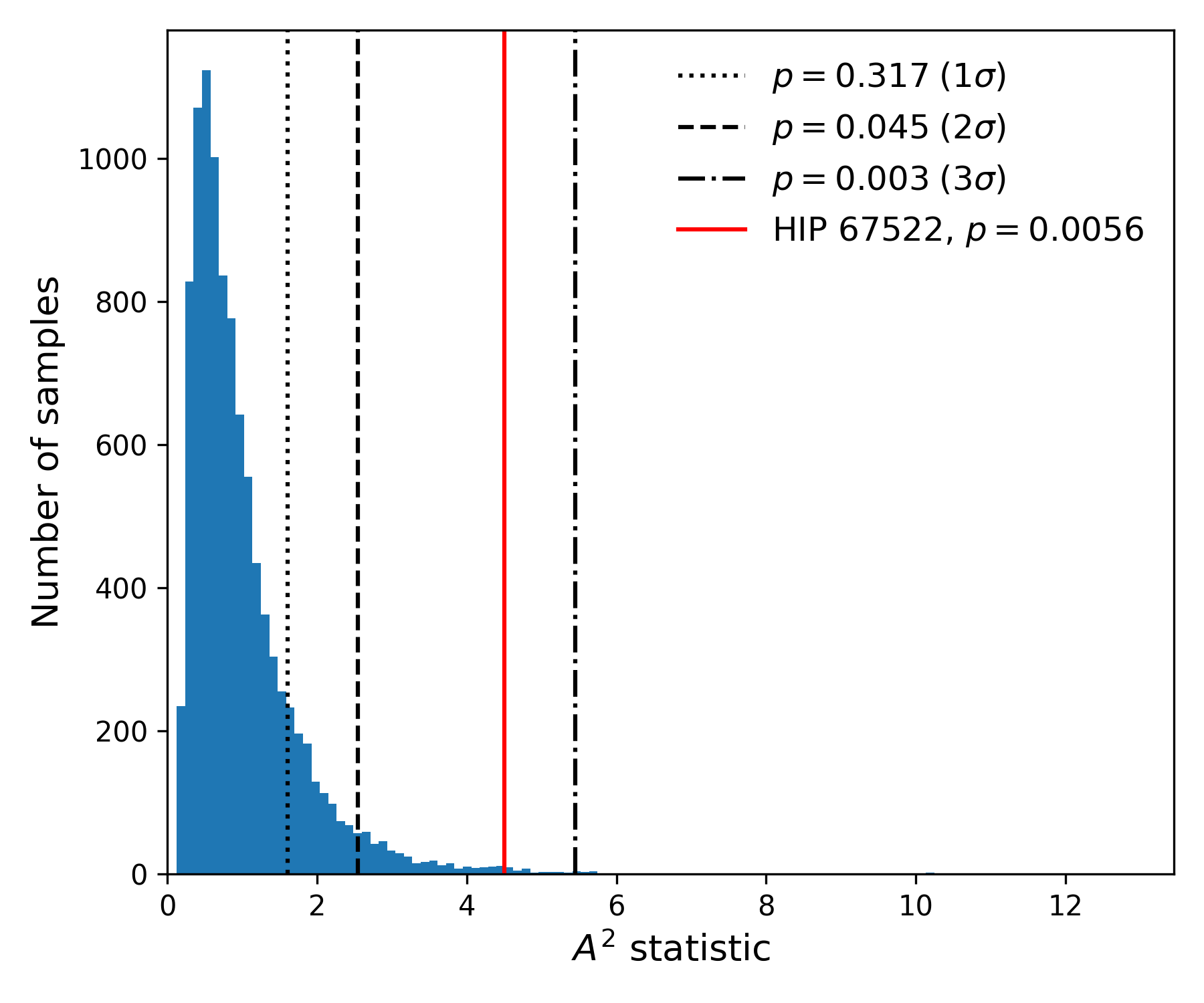}
\caption{The distribution of 10000 samples of the $A^2$ statistic for HIP 67522. A value of $A^2$ is calculated by drawing a fixed number of flares \textit{at random orbital phases}. The fixed number has to be the same as the number of the flares that were observed in reality. Then, one takes the orbital phases of these observed flares, and compares their $A^2$ value to that of the distribution. An observed $A^2$ value in the tails of the distribution means that flares do not occur \textit{at random orbital phases}, but instead cluster in phase with the planetary orbit, suggesting that the planet triggers at least some of those flares. Here, we show the $A^2$ values that would correspond to $1$, $2$, and $3\sigma$ detections of clustering in orbital phase as black vertical lines, along with the measured $A^2$ value for HIP 67522 as red vertical line.}
    \label{fig:ad_dist_hip}
        \end{centering}
\end{figure*}

\end{document}

%% file: output/table_lit_vals.tex
\begin{tabular}{lllllllllll}
\hline
ID & $P_{\rm rot}$ & $P_{\rm orb}$ & $R_{*}$ & SpT & $R_{\rm p}$ & $a$ & $a/R_*$ & $e$ & log$_{10} L_{*}$ & obs. time \\

 & [d] & [d] & [R$_\odot$] &  & [R$_J$] & [$10^{-2}$ au] &  &  & [L$_\odot$] & [d] \\ \hline
Kepler-1558 & $12.16 [4]$ (3) & $3.50470 [2]$ (40) & $0.79 [4]$ (40) & K2.5 V (33, 43) & $0.061^{0.005}_{ -0.004}$ (40) & $4 [1]$ (49) & $11 [3]$ & - (49) & $-0.474^{0.007}_{ -0.008}$ (17) & 74.3 \\
L 98-59 & $81 [5]$ (14) & $2.253114 [1]$ (14) & $0.30 [3]$ (14) & M3 V (14) & $0.076^{0.005}_{ -0.004}$ (14) & $2.2 [2]$ (14) & $16 [2]$ & $0.10^{0.12}_{ -0.04}$ (14) & $-1.95^{0.02}_{ -0.02}$ (14) & 335.5 \\
Kepler-367 & $15.41 [6]$ (3) & $37.8157 [3]$ (45) & $0.69 [4]$ (45) & K3 V (33, 43) & $0.12^{0.01}_{ -0.01}$ (45) & $20 [5]$ (45) & $60 [16]$ & - (49) & $-0.62^{0.02}_{ -0.02}$ (46) & 438.3 \\
Kepler-42 & $68 [2]$ (36) & $0.4532851 [10]$ (41) & $0.17 [4]$ (41) & M3 V (25) & $0.07^{0.02}_{ -0.02}$ (41) & $0.60 [6]$ (41) & $8 [2]$ & $0.0^{0.1}_{ 0.0}$ (32) & $-2.6^{0.3}_{ -0.3}$ (41) & 828.5 \\
Kepler-55 & $13.720 [1]$ (36) & $2.211099 [4]$ (45) & $0.62 [3]$ (45) & K6 V (19, 43) & $0.142^{0.008}_{ -0.008}$ (45) & $2.9 [7]$ (45) & $10 [3]$ & - (49) & $-0.75^{0.02}_{ -0.02}$ (17) & 882.9 \\
Kepler-862 & $26.7 [3]$ (37) & $3.148665 [7]$ (40) & $0.84 [7]$ (40) & K0 V (8, 43) & $0.20^{0.04}_{ -0.01}$ (40) & $3.9 [10]$ (49) & $10 [3]$ & - (49) & $-0.19^{0.03}_{ -0.03}$ (17) & 25.4 \\
GJ 3082 & $6.6 [-]$ (24) & $11.95 [2]$ (16) & $0.47 [1]$ (46) & M1 V (20, 43) & $0.25^{0.11}_{ -0.07}$ (12) & $8 [2]$ (16) & $36 [8]$ & $0.2^{0.1}_{ -0.1}$ (16) & $-1.5^{0.1}_{ -0.1}$ (46) & 47.1 \\
GJ 674 & $33.3 [-]$ (28) & $4.694 [7]$ (9) & $0.36 [1]$ (46) & M2.5 V (9) & $0.30^{0.11}_{ -0.08}$ (12) & $3.9 [8]$ (9) & $23 [5]$ & $0.20^{0.02}_{ -0.02}$ (9) & $-1.760^{0.008}_{ -0.007}$ (17) & 47.6 \\
GJ 3323 & $33.0 [-]$ (5) & $5.3636 [7]$ (4) & $0.12 [3]$ (4) & M4 V (4) & $0.11^{0.03}_{ -0.02}$ (12) & $3.3 [8]$ (4) & $60 [20]$ & $0.2^{0.1}_{ -0.1}$ (4) & $-3^{1}_{ -1}$ (4) & 48.0 \\
TAP 26 & $0.7 [-]$ (23) & $10.8 [1]$ (53) & $1.2 [2]$ (53) & K4 V (26) & $1.2^{0.2}_{ -0.2}$ (12) & $10 [2]$ (53) & $18 [4]$ & $0.2^{0.1}_{ -0.1}$ (53) & $-0.2^{0.1}_{ -0.1}$ (53) & 44.2 \\
TRAPPIST-1 & $3.3 [1]$ (30) & $1.510826 [6]$ (1) & $0.119 [1]$ (1) & M8 V (22) & $0.100^{0.001}_{ -0.001}$ (1) & $1.15 [1]$ (1) & $20.8 [3]$ & $0.006^{0.003}_{ -0.003}$  & $-3.26^{0.01}_{ -0.01}$ (1) & 73.1 \\
YZ Cet & $68.3 [3]$ (48) & $2.02087 [8]$ (48) & $0.16 [1]$ (48) & M4.5 V (48) & $0.082^{0.010}_{ -0.009}$ (12) & $1.63 [10]$ (48) & $22 [2]$ & $0.06^{0.06}_{ -0.04}$ (48) & $-2.659^{0.008}_{ -0.008}$ (48) & 40.5 \\
KOI-12 & $1.2 [1]$ (36) & $17.855182 [7]$ (35) & $1.40 [7]$ (35) & F5 IV (18) & $1.23^{0.06}_{ -0.05}$ (35) & $15 [5]$ (35) & $24 [8]$ & $0.34^{0.08}_{ -0.07}$ (35) & $0.63^{0.03}_{ -0.03}$ (46) & 1375.6 \\
HAT-P-11 & $28 [3]$ (3) & $4.887802 [7]$ (6) & $0.68 [1]$ (52) & K4 V (6) & $0.389^{0.005}_{ -0.005}$ (52) & $5 [1]$ (52) & $17 [4]$ & $0.22^{0.03}_{ -0.03}$ (52) & $-0.57^{0.01}_{ -0.02}$ (46) & 1043.7 \\
K2-354\footnote{We note that we found K2-354 under EPIC 211314705 and K2-329(!) in \citet{bouma2020cluster}, who refer to the detection paper \citep{pope2016transiting}, and also under TIC 468989066.} & $10.4 [8]$ (13) & $3.7933 [3]$ (13) & $0.41 [1]$ (13) & M1 V (13, 43) & $0.139^{0.007}_{ -0.007}$ (13) & $3.6 [9]$ (13) & $19 [5]$ & -  & $-1.6^{0.1}_{ -0.1}$ (46) & 90.1 \\
Kepler-249 & $36.4 [3]$ (37) & $3.30654 [1]$ (45) & $0.48 [2]$ (45) & M2 V (42) & $0.097^{0.004}_{ -0.004}$ (45) & $3.5 [9]$ (45) & $16 [4]$ & - (49) & $-1.50^{0.07}_{ -0.09}$ (31) & 727.7 \\
GJ 687 & $61.7 [-]$ (10) & $38.142 [7]$ (15) & $0.42 [1]$ (10) & M3 V (10) & $0.4^{0.2}_{ -0.1}$ (12) & $16 [3]$ (15) & $80 [14]$ & $0.17^{0.05}_{ -0.05}$ (15) & $-1.672^{0.005}_{ -0.005}$ (10) & 437.9 \\
Kepler-705 & $20 [1]$ (50) & $56.0561 [2]$ (40) & $0.51 [1]$ (40) & M2 V (42) & $0.188^{0.009}_{ -0.007}$ (40) & $23 [6]$ (50) & $100 [25]$ & - (49) & $-1.39^{0.05}_{ -0.06}$ (50) & 452.9 \\
Proxima Cen & $83.0 [-]$ (2) & $11.186 [2]$ (2) & $0.14 [2]$ (2) & M5.5 V (2) & $0.096^{0.012}_{ -0.010}$ (12) & $5 [2]$ (2) & $70 [28]$ & $0.3^{-}_{ -}$ (2) & $-2.81^{0.02}_{ -0.02}$ (2) & 72.2 \\
Kepler-1646 & $3.514 [6]$ (38) & $4.48558 [1]$ (40) & $0.26 [5]$ (40) & M4 V (25) & $0.11^{0.03}_{ -0.02}$ (40) & $2.9 [7]$ (49) & $24 [8]$ & - (49) & $-1.8^{0.1}_{ -0.1}$ (46) & 76.0 \\
Kepler-396 & $13.31 [5]$ (3) & $42.99292 [2]$ (7) & $1.1 [4]$ (51) & G5 V (33, 43) & $0.31^{0.11}_{ -0.06}$ (51) & $23 [6]$ (49) & $50 [18]$ & - (49) & $-0.14^{0.01}_{ -0.01}$ (46) & 546.3 \\
K2-25 & $1.878 [5]$ (47) & $3.4845641 [6]$ (47) & $0.29 [1]$ (47) & M4.5 V (47) & $0.31^{0.01}_{ -0.01}$ (47) & $3 [1]$ (47) & $21 [9]$ & $0.43^{0.05}_{ -0.05}$ (47) & $-2.09^{0.01}_{ -0.02}$ (47) & 21.8 \\
AU Mic & $4.857 [4]$ (54) & $8.463000 [2]$ (54) & $0.75 [3]$ (21) & M1 V (29) & $0.36^{0.01}_{ -0.01}$ (34) & $6 [1]$ (11) & $18 [4]$ & $0.19^{0.04}_{ -0.04}$ (11) & $-1.0^{0.1}_{ -0.1}$ (54) & 48.4 \\
TOI-540 & $0.7261 [4]$ (39) & $1.239149 [2]$ (39) & $0.19 [1]$ (39) & M4 V (39) & $0.081^{0.005}_{ -0.005}$ (39) & $1.2 [3]$ (39) & $14 [4]$ & - (39) & $-2.46^{0.03}_{ -0.03}$ (39) & 111.2 \\
HIP 67522 & $1.42 [2]$ (44) & $6.95950 [2]$ (44) & $1.38 [6]$ (44) & G1 V (27) & $0.90^{0.04}_{ -0.04}$ (44) & $7.5 [4]$ (44) & $11.7 [9]$ & $0.06^{0.19}_{ -0.05}$ (44) & $0.24^{0.02}_{ -0.02}$ (44) & 92.1 \\
\hline
\end{tabular}

%% file: output/lit_table_bibstring.tex
(1) \citet{agol2021refining}, (2) \citet{anglada-escude2016terrestrial}, (3) \citet{angus2018inferring}, (4) \citet{astudillo-defru2017harps}, (5) \citet{astudillo-defru2017magnetic}, (6) \citet{bakos2010hatp11b}, (7) \citet{battley2021revisiting}, (8) \citet{berger2018revised}, (9) \citet{bonfils2007harps}, (10) \citet{burt2014lickcarnegie}, (11) \citet{cale2021diving}, (12) \citet{chen2017probabilistic}, (13) \citet{deleon202137}, (14) \citet{demangeon2021warm}, (15) \citet{feng2020search}, (16) \citet{feng2020searcha}, (17) \citet{fouesneau2022gaia}, (18) \citet{frasca2016activity}, (19) \citet{furlan2018kepler}, (20) \citet{gaidos2014trumpeting}, (21) \citet{gilbert2022flares}, (22) \citet{gillon2016temperate}, (23) \citet{grankin2013magnetically}, (24) \citet{gunther2020stellar}, (25) \citet{hardegree-ullman2019kepler}, (26) \citet{herczeg2014optical}, (27) \citet{houk1978michigan}, (28) \citet{kiraga2007agerotationactivity}, (29) \citet{klein2021investigating}, (30) \citet{luger2017sevenplanet}, (31) \citet{mann2013spectrothermometry}, (32) \citet{mann2017gold}, (33) \citet{martinez2019spectroscopic}, (34) \citet{martioli2021new}, (35) \citet{masuda2017eccentric}, (36) \citet{mazeh2015photometric}, (37) \citet{mcquillan2013stellar}, (38) \citet{mcquillan2014rotation}, (39) \citet{ment2021toi}, (40) \citet{morton2016false}, (41) \citet{muirhead2012characterizing}, (42) \citet{muirhead2014characterizing}, (43) \citet{pecaut2013intrinsic}, (44) \citet{rizzuto2020tess}, (45) \citet{rowe2014validation}, (46) \citet{stassun2019revised}, (47) \citet{stefansson2020habitable}, (48) \citet{stock2020carmenes}, (49) \citet{thompson2018planetary}, (50) \citet{torres2017validation}, (51) \citet{xie2014transit}, (52) \citet{yee2018hatp11}, (53) \citet{yu2017hot}, (54) \citet{zicher2022one}.

%% file: output/PAPER_total_number_of_flares.txt
1169

%% file: output/PAPER_total_number_of_systems_with_flares.txt
92

%% file: output/flare_table.tex
\begin{tabular}{lcccccccc}
\hline
TIC & ID & mission & Qua./Sec. & $t_s$ [BKJD/BTJD] & $t_f$ [BKJD/BTJD] & orb. phase & $a$ & $ED$ [s] \\
\hline
169461816 & KOI-12 & Kepler & 0 & $124.0381$ & $124.0422$ & $0.7366002 [5]$ & $0.0018$ & $0.42 [3]$ \\
299096355 & Kepler-16 & Kepler & 2 & $221.7214$ & $221.7350$ & $0.31198 [9]$ & $0.0045$ & $2.38 [8]$ \\
139106731 & Kepler-908 & Kepler & 4 & $431.9578$ & $431.9659$ & $0.5670 [3]$ & $0.0063$ & $2.00 [4]$ \\
139106731 & Kepler-908 & Kepler & 4 & $433.3548$ & $433.3568$ & $0.6091 [3]$ & $0.0012$ & $0.18 [2]$ \\
138297607 & Kepler-636 & Kepler & 8 & $745.8702$ & $745.8730$ & $0.01785 [1]$ & $0.0033$ & $0.62 [9]$ \\
138297607 & Kepler-636 & Kepler & 8 & $749.3248$ & $749.3275$ & $0.23268 [1]$ & $0.0031$ & $0.61 [10]$ \\
138297607 & Kepler-636 & Kepler & 8 & $749.4651$ & $749.4678$ & $0.24141 [1]$ & $0.0046$ & $0.87 [8]$ \\
26417717 & Kepler-808 & Kepler & 9 & $860.1291$ & $860.1441$ & $0.2421 [6]$ & $0.0110$ & $7.1 [3]$ \\
26417717 & Kepler-808 & Kepler & 10 & $911.4007$ & $911.4088$ & $0.4538 [7]$ & $0.0053$ & $2.7 [2]$ \\
28230919 & HAT-P-11 & Kepler & 10 & $930.8736$ & $930.8811$ & $0.91263 [5]$ & $0.0018$ & $0.51 [2]$ \\
28230919 & HAT-P-11 & Kepler & 10 & $935.8758$ & $935.8874$ & $0.93604 [6]$ & $0.0014$ & $0.80 [4]$ \\
399954349(c) & Kepler-411(c) & Kepler & 11 & $1002.9499$ & $1002.9547$ & $0.1 [nan]$ & $0.0059$ & $1.06 [5]$ \\
399954349(c) & Kepler-411(c) & Kepler & 11 & $1003.9879$ & $1003.9900$ & $0.5 [nan]$ & $0.0027$ & $0.37 [4]$ \\
399954349(c) & Kepler-411(c) & Kepler & 11 & $1007.9104$ & $1007.9138$ & $0.8 [nan]$ & $0.0042$ & $0.78 [4]$ \\
399954349(c) & Kepler-411(c) & Kepler & 11 & $1008.0412$ & $1008.0446$ & $0.8 [nan]$ & $0.0034$ & $0.67 [5]$ \\
417676622 & Kepler-68 & Kepler & 11 & $1008.4671$ & $1008.4712$ & $0.5926 [1]$ & $0.0008$ & $0.23 [2]$ \\
26417717 & Kepler-808 & Kepler & 11 & $1010.6739$ & $1010.6766$ & $0.6979 [8]$ & $0.0042$ & $0.7 [1]$ \\
399954349(c) & Kepler-411(c) & Kepler & 11 & $1010.8759$ & $1010.8786$ & $0.8 [nan]$ & $0.0046$ & $0.72 [4]$ \\
399954349(c) & Kepler-411(c) & Kepler & 11 & $1012.3975$ & $1012.4023$ & $0.3 [nan]$ & $0.0032$ & $0.84 [6]$ \\
399954349(c) & Kepler-411(c) & Kepler & 11 & $1015.7363$ & $1015.7390$ & $0.4 [nan]$ & $0.0020$ & $0.40 [4]$ \\
\hline

\end{tabular}

%% file: output/table_der_vals.tex
\begin{tabular}{lllllllll}
\hline
ID & $R$o & $B$ & $v_{\mathrm{rel}}$ & log$_{10} P_{\rm spi,sb}$ & log$_{10} P_{\rm spi,sb0}$ & log$_{10} P_{\rm spi,aw}$ & log$_{10} P_{\rm spi,aw0}$ & $p$-value \\

 &  & [G] & [km s$^{-1}$] & [erg s$^{-1}$] & [erg s$^{-1}$] & [erg s$^{-1}$] & [erg s$^{-1}$] &  \\ \hline
Kepler-1558 & $0.573^{0.007}_{ -0.007}$ & $400^{30}_{-30}$ & $90 [23]$ & $24.1^{0.8}_{ -0.7}$ & $25.8^{0.8}_{ -0.7}$ & $21.8^{0.6}_{ -0.6}$ & $21.3^{0.9}_{ -0.8}$ & $0.93 [1]$ \\
L 98-59 & $0.70^{0.06}_{ -0.06}$ & $310^{50}_{-40}$ & $100 [11]$ & $23.6^{0.5}_{ -0.4}$ & $25.3^{0.6}_{ -0.5}$ & $21.7^{0.3}_{ -0.3}$ & $21.0^{0.6}_{ -0.5}$ & $0.92 [2]$ \\
Kepler-367 & $0.62^{0.02}_{ -0.02}$ & $370^{40}_{-30}$ & $-80 [21]$ & $21.8^{0.8}_{ -0.7}$ & $23.5^{0.8}_{ -0.8}$ & $20.8^{0.6}_{ -0.6}$ & $18.8^{0.9}_{ -0.9}$ & $0.87 [2]$ \\
Kepler-42 & $0.27^{0.12}_{ -0.08}$ & $1000^{900}_{-400}$ & $140 [14]$ & $26^{1}_{ -1}$ & $28^{1}_{ -1}$ & $22.7^{0.7}_{ -0.8}$ & $23^{1}_{ -1}$ & $0.84 [3]$ \\
Kepler-55 & $0.47^{0.01}_{ -0.01}$ & $510^{60}_{-50}$ & $120 [30]$ & $25.3^{0.8}_{ -0.7}$ & $27.1^{0.8}_{ -0.7}$ & $22.9^{0.5}_{ -0.6}$ & $22.6^{0.9}_{ -0.8}$ & $0.84 [2]$ \\
Kepler-862 & $1.73^{0.09}_{ -0.08}$ & $100^{10}_{-10}$ & $120 [30]$ & $24.6^{0.9}_{ -0.7}$ & $25.9^{0.9}_{ -0.8}$ & $22.9^{0.7}_{ -0.6}$ & $22.2^{1.0}_{ -0.9}$ & $0.77 [3]$ \\
GJ 3082 & $0.10^{0.02}_{ -0.02}$ & $2600^{100}_{-100}$ & $-60 [19]$ & $24.3^{0.9}_{ -0.9}$ & $26.6^{0.9}_{ -0.9}$ & $21.9^{0.8}_{ -0.8}$ & $21^{1}_{ -1}$ & $0.7 [2]$ \\
GJ 674 & $0.36^{0.04}_{ -0.04}$ & $700^{200}_{-100}$ & $80 [16]$ & $24.6^{0.9}_{ -0.9}$ & $26.5^{1.0}_{ -0.9}$ & $22.5^{0.7}_{ -0.7}$ & $21.6^{1.0}_{ -0.9}$ & $0.7 [2]$ \\
GJ 3323 & $0.1^{0.1}_{ -0.1}$ & $2000^{29000}_{-2000}$ & $60 [13]$ & $23^{3}_{ -2}$ & $25^{3}_{ -2}$ & $20.7^{1.2}_{ -1.0}$ & $19^{2}_{ -2}$ & $0.6 [2]$ \\
TAP 26 & $0.043^{0.010}_{ -0.009}$ & $2900^{200}_{-200}$ & $-1400 [265]$ & $28.3^{0.8}_{ -0.8}$ & $30.6^{0.8}_{ -0.8}$ & $26.6^{0.6}_{ -0.6}$ & $26.4^{0.9}_{ -0.9}$ & $0.6 [2]$ \\
TRAPPIST-1 & $0.0063^{0.0004}_{ -0.0004}$ & $3600^{600}_{-500}$ & $45 [2]$ & $24.5^{0.1}_{ -0.1}$ & $26.8^{0.2}_{ -0.2}$ & $21.37^{0.08}_{ -0.08}$ & $21.1^{0.1}_{ -0.1}$ & $0.55 [5]$ \\
YZ Cet & $0.260^{0.003}_{ -0.003}$ & $1100^{200}_{-200}$ & $85 [5]$ & $23.8^{0.4}_{ -0.4}$ & $25.8^{0.5}_{ -0.5}$ & $21.5^{0.3}_{ -0.3}$ & $20.8^{0.4}_{ -0.4}$ & $0.5 [3]$ \\
KOI-12 & $0.21^{0.03}_{ -0.03}$ & $1500^{700}_{-400}$ & $-1300 [458]$ & $27.4^{1.1}_{ -1.0}$ & $29^{1}_{ -1}$ & $26.2^{0.8}_{ -0.8}$ & $26^{1}_{ -1}$ & $0.54 [1]$ \\
HAT-P-11 & $1.2^{0.1}_{ -0.1}$ & $160^{30}_{-20}$ & $100 [21]$ & $24.6^{0.6}_{ -0.6}$ & $26.0^{0.7}_{ -0.6}$ & $23.0^{0.4}_{ -0.4}$ & $21.9^{0.7}_{ -0.7}$ & $0.534 [8]$ \\
K2-354 & $0.13^{0.03}_{ -0.03}$ & $2600^{1700}_{-900}$ & $70 [17]$ & $24.9^{0.9}_{ -0.8}$ & $27.2^{1.1}_{ -0.9}$ & $22.0^{0.6}_{ -0.6}$ & $21.7^{1.0}_{ -0.9}$ & $0.51 [2]$ \\
Kepler-249 & $0.53^{0.05}_{ -0.06}$ & $440^{110}_{-70}$ & $100 [26]$ & $24.1^{0.8}_{ -0.7}$ & $25.8^{0.9}_{ -0.7}$ & $22.0^{0.5}_{ -0.5}$ & $21.3^{0.9}_{ -0.8}$ & $0.50 [2]$ \\
GJ 687 & $0.73^{0.08}_{ -0.08}$ & $300^{60}_{-40}$ & $18 [4]$ & $21.5^{0.8}_{ -0.8}$ & $23.1^{0.9}_{ -0.8}$ & $20.1^{0.7}_{ -0.7}$ & $17.9^{0.9}_{ -0.9}$ & $0.4 [2]$ \\
Kepler-705 & $0.33^{0.04}_{ -0.04}$ & $800^{300}_{-200}$ & $-80 [21]$ & $21.9^{0.8}_{ -0.7}$ & $23.8^{0.9}_{ -0.8}$ & $20.9^{0.5}_{ -0.6}$ & $18.8^{0.9}_{ -0.9}$ & $0.43 [2]$ \\
Proxima Cen & $0.27^{0.03}_{ -0.03}$ & $1100^{400}_{-200}$ & $40 [14]$ & $22^{1}_{ -1}$ & $24^{1}_{ -1}$ & $20.0^{0.9}_{ -0.9}$ & $18^{1}_{ -1}$ & $0.3 [2]$ \\
Kepler-1646 & $0.037^{0.004}_{ -0.005}$ & $2900^{300}_{-200}$ & $-20 [5]$ & $24^{1}_{ -1}$ & $26^{1}_{ -1}$ & $20.6^{0.8}_{ -0.8}$ & $20^{1}_{ -1}$ & $0.27 [1]$ \\
Kepler-396 & $0.93^{0.02}_{ -0.02}$ & $219^{8}_{ -6}$ & $-130 [33]$ & $23^{1}_{ -1}$ & $25^{1}_{ -1}$ & $22.2^{1.0}_{ -0.8}$ & $20^{2}_{ -1}$ & $0.22 [2]$ \\
K2-25 & $0.0138^{0.0003}_{ -0.0003}$ & $3300^{400}_{-400}$ & $-80 [33]$ & $25.6^{1.2}_{ -1.0}$ & $28^{1}_{ -1}$ & $22.8^{0.9}_{ -0.9}$ & $22^{1}_{ -1}$ & $0.18 [3]$ \\
AU Mic & $0.12^{0.01}_{ -0.01}$ & $2600^{100}_{-100}$ & $-60 [11]$ & $25.7^{0.5}_{ -0.5}$ & $28.0^{0.6}_{ -0.5}$ & $22.8^{0.4}_{ -0.4}$ & $22.5^{0.6}_{ -0.6}$ & $0.12 [2]$ \\
TOI-540 & $0.0035^{0.0001}_{ -0.0001}$ & $3800^{700}_{-600}$ & $-80 [19]$ & $25.2^{0.8}_{ -0.7}$ & $27.6^{0.8}_{ -0.8}$ & $22.0^{0.6}_{ -0.6}$ & $22.1^{0.9}_{ -0.9}$ & $0.084 [9]$ \\
HIP 67522 & $0.153^{0.006}_{ -0.006}$ & $2100^{600}_{-400}$ & $-460 [28]$ & $28.0^{0.4}_{ -0.4}$ & $30.2^{0.4}_{ -0.4}$ & $25.7^{0.2}_{ -0.2}$ & $25.8^{0.4}_{ -0.4}$ & $0.0056 [5]$ \\
\hline
\end{tabular}

%% file: output/multiples_string.tex
Kepler-1627~\citep{kraus2016impact}, Kepler-636~\citep{baranec2016roboao,ziegler2018measuring,kraus2016impact}, Kepler-808~\citep{baranec2016roboao,ziegler2018measuring, kraus2016impact}, Kepler-155~\citep{baranec2016roboao,ziegler2018measuring}, DS Tuc A~\citep{newton2019tess}, TOI-837~\citep{bouma2020cluster}, Kepler-1651~\citep{kraus2016impact}, HD 41004 B~\citep{zucker2004multiorder}, LTT 1445 A~\citep{winters2019three}.

%% file: output/table_tidal.tex
\begin{tabular}{lllllllcc}
\hline
ID & $M_*$ [$M_\odot$] & $M_{\mathrm{p}} (\sin i)$ [$M_\oplus$] & log$_{10} 10^{-8} \Delta g / g$ & log$_{10} \tau_{\rm tide}$ [yr] & $10^{-18} \frac{\partial L_{\rm conv}}{\partial t}$ $\left[M_\odot \left(\frac{km}{s}\right)^2\right]$ & $p$-value & ref. $M_*$ & ref. $M_{\rm p}$ \\
\hline
TAP 26 & $1.0^{0.1}_{ -0.1}$ & $530^{100}_{-100}$ & $1.8^{0.5}_{ -0.5}$ & $3.8^{0.6}_{ -0.5}$ & $-3000^{-20000}_{3000}$ & 0.93 [0.04] & 23 & 23 \\
Kepler-396 & $0.85^{0.13}_{ -0.06}$ & $75.7^{11.8}_{ -5.7}$ & $-0.3^{0.9}_{ -0.7}$ & $7.9^{0.9}_{ -1.4}$ & $-0.2^{-6.4}_{ 0.2}$ & 0.88 [0.10] & 21 & 21 \\
K2-25 & $0.26^{0.01}_{ -0.01}$ & $24.5^{5.7}_{ -5.2}$ & $0.8^{0.9}_{ -0.6}$ & $5.8^{0.8}_{ -1.3}$ & $-8.6^{-434.2}_{ 8.1}$ & 0.84 [0.07] & 19 & 19 \\
Kepler-1646 & $0.24^{0.06}_{ -0.06}$ & $2.0^{2.1}_{ -1.0}$ & $-0.4^{1.0}_{ -1.0}$ & $8.2^{0.3}_{ -0.4}$ & $-0.03^{-1.58}_{ 0.03}$ & 0.78 [0.06] & 14 & 6 \\
Kepler-862 & $0.88^{0.04}_{ -0.05}$ & $6.5^{5.6}_{ -2.8}$ & $0.8^{0.8}_{ -0.7}$ & $6.0^{0.2}_{ -0.4}$ & $20^{550}_{-20}$ & 0.73 [0.04] & 14 & 6 \\
Kepler-705 & $0.53^{0.02}_{ -0.02}$ & $5.5^{4.1}_{ -2.2}$ & $-2.2^{0.7}_{ -0.6}$ & $11.6^{0.2}_{ -0.3}$ & $-0.00003^{-0.00047}_{ 0.00002}$ & 0.69 [0.03] & 14 & 6 \\
Kepler-367 & $0.68^{0.02}_{ -0.02}$ & $2.2^{1.8}_{ -0.9}$ & $-2.1^{0.7}_{ -0.6}$ & $11.5^{0.2}_{ -0.4}$ & $-0.00004^{-0.00107}_{ 0.00004}$ & 0.58 [0.05] & 4 & 6 \\
HIP 67522 & $1.22^{0.05}_{ -0.05}$ & $1589.5^{0.0}_{ -1508.3}$ & $2.8^{0.2}_{ -1.4}$ & $1.9^{2.3}_{ 0.2}$ & $-300000^{-200000}_{300000}$ & 0.49 [0.09] & 16 & 6 \\
Kepler-249 & $0.51^{0.02}_{ -0.02}$ & $1.3^{0.9}_{ -0.5}$ & $-0.3^{0.7}_{ -0.6}$ & $8.1^{0.2}_{ -0.5}$ & $0.10^{1.71}_{ -0.09}$ & 0.47 [0.03] & 18 & 6 \\
AU Mic & $0.50^{0.03}_{ -0.03}$ & $20.1^{1.6}_{ -1.7}$ & $0.7^{0.4}_{ -0.3}$ & $6.1^{0.4}_{ -0.5}$ & $-5.2^{-20.1}_{ 3.9}$ & 0.44 [0.15] & 5 & 5 \\
Kepler-55 & $0.63^{0.02}_{ -0.02}$ & $3.5^{2.5}_{ -1.3}$ & $0.7^{0.7}_{ -0.6}$ & $6.2^{0.2}_{ -0.5}$ & $7.6^{150.8}_{ -6.9}$ & 0.43 [0.03] & 4 & 6 \\
GJ 3323 & $0.198^{0.004}_{ -0.004}$ & $2.0^{0.3}_{ -0.3}$ & $-1.5^{0.7}_{ -0.7}$ & $10.3^{1.2}_{ -1.1}$ & $0.0003^{0.0056}_{ -0.0003}$ & 0.41 [0.09] & 11 & 2 \\
HAT-P-11 & $0.81^{0.02}_{ -0.03}$ & $26.7^{2.2}_{ -2.2}$ & $0.7^{0.4}_{ -0.3}$ & $6.0^{0.5}_{ -0.6}$ & $20^{90}_{-10}$ & 0.40 [0.01] & 22 & 17 \\
GJ 3082 & $0.487^{0.010}_{ -0.010}$ & $8.2^{1.7}_{ -1.7}$ & $-0.6^{0.5}_{ -0.4}$ & $8.6^{0.4}_{ -0.5}$ & $-0.02^{-0.15}_{ 0.02}$ & 0.39 [0.06] & 11 & 10 \\
YZ Cet & $0.14^{0.01}_{ -0.01}$ & $0.70^{0.09}_{ -0.08}$ & $-0.5^{0.2}_{ -0.2}$ & $8.4^{0.2}_{ -0.2}$ & $0.011^{0.016}_{ -0.006}$ & 0.39 [0.07] & 20 & 20 \\
Proxima Cen & $0.12^{0.01}_{ -0.01}$ & $1.3^{0.2}_{ -0.2}$ & $-1.8^{0.8}_{ -0.7}$ & $10.9^{1.0}_{ -1.3}$ & $0.00003^{0.00087}_{ -0.00002}$ & 0.38 [0.27] & 1 & 1 \\
K2-354 & $0.43^{0.01}_{ -0.01}$ & $3.5^{2.4}_{ -1.2}$ & $-0.08^{0.66}_{ -0.54}$ & $7.6^{0.2}_{ -0.4}$ & $0.2^{4.0}_{ -0.2}$ & 0.34 [0.02] & 7 & 6 \\
GJ 674 & $0.376^{0.007}_{ -0.007}$ & $11.1^{0.3}_{ -0.3}$ & $0.2^{0.4}_{ -0.3}$ & $7.0^{0.5}_{ -0.6}$ & $0.8^{3.0}_{ -0.6}$ & 0.31 [0.13] & 11 & 3 \\
L 98-59 & $0.27^{0.03}_{ -0.03}$ & $0.4^{0.2}_{ -0.1}$ & $-0.6^{0.5}_{ -0.5}$ & $8.57^{0.04}_{ -0.03}$ & $0.01^{0.07}_{ -0.01}$ & 0.25 [0.02] & 8 & 8 \\
GJ 687 & $0.40^{0.02}_{ -0.02}$ & $17.2^{1.0}_{ -1.0}$ & $-1.3^{0.3}_{ -0.3}$ & $10.0^{0.4}_{ -0.5}$ & $0.0008^{0.0023}_{ -0.0006}$ & 0.23 [0.14] & 9 & 9 \\
Kepler-1558 & $0.83^{0.04}_{ -0.04}$ & $0.37^{0.07}_{ -0.07}$ & $-0.6^{0.6}_{ -0.5}$ & $8.8^{0.5}_{ -0.7}$ & $0.03^{0.27}_{ -0.03}$ & 0.20 [0.02] & 14 & 6 \\
TOI-540 & $0.16^{0.01}_{ -0.01}$ & $0.7^{0.5}_{ -0.2}$ & $0.10^{0.72}_{ -0.59}$ & $7.3^{0.2}_{ -0.5}$ & $-0.1^{-2.8}_{ 0.1}$ & 0.18 [0.01] & 13 & 6 \\
Kepler-42 & $0.13^{0.05}_{ -0.05}$ & $0.5^{1.6}_{ -0.4}$ & $0.9^{1.3}_{ -1.2}$ & $5.9^{0.8}_{ -0.8}$ & $2.5^{270.9}_{ -2.4}$ & 0.10 [0.01] & 15 & 15 \\
KOI-12 & $1.5^{0.1}_{ -0.1}$ & $300^{1100}_{-300}$ & $1.1^{1.3}_{ -1.0}$ & $5.24^{-0.03}_{ -0.43}$ & $-200^{-51800}_{200}$ & 0.09 [0.03] & 12 & 12 \\
\hline
\end{tabular}

%% file: output/table_tidal_bibstring.tex
(1) \citet{anglada-escude2016terrestrial}, (2) \citet{astudillo-defru2017harps}, (3) \citet{bonfils2007harps}, (4) \citet{burke2014planetary}, (5) \citet{cale2021diving}, (6) \citet{chen2017probabilistic}, (7) \citet{deleon202137}, (8) \citet{demangeon2021warm}, (9) \citet{feng2020search}, (10) \citet{feng2020searcha}, (11) \citet{mann2015how}, (12) \citet{masuda2017eccentric}, (13) \citet{ment2021toi}, (14) \citet{morton2016false}, (15) \citet{muirhead2012characterizing}, (16) \citet{rizzuto2020tess}, (17) \citet{southworth2011homogeneous}, (18) \citet{stassun2019revised}, (19) \citet{stefansson2020habitable}, (20) \citet{stock2020carmenes}, (21) \citet{xie2014transit}, (22) \citet{yee2018hatp11}, (23) \citet{yu2017hot}, 